\documentclass[a4paper,12pt]{article}
\usepackage{amssymb}
\usepackage{amsmath}
\usepackage{epsfig}
\usepackage{subfigure}
\usepackage{graphics}

\usepackage{latexsym}
\usepackage{rotating}
\usepackage{titlesec}

\title{Nonlocal Nonlinear Schr\"{o}dinger Equations and Their Soliton Solutions}

\author{Metin G\"{u}rses \thanks{gurses@fen.bilkent.edu.tr}\\
{\small Department of Mathematics, Faculty of Science}\\
{\small Bilkent University, 06800 Ankara - Turkey}\\
Asl{\i} Pekcan \thanks{Email:aslipekcan@hacettepe.edu.tr} \\
{\small Department of Mathematics, Faculty of Science} \\
{\small Hacettepe University, 06800 Ankara - Turkey}
}

\date{\nonumber}
\begin{document}
\maketitle
\date{\nonumber}
\newtheorem{thm}{Theorem}[section]
\newtheorem{Le}{Lemma}[section]
\newtheorem{defi}{Definition}[section]
\newtheorem{ex}{Example}[section]
\newtheorem{pro}{Proposition}[section]
\baselineskip 17pt

\numberwithin{equation}{section}

\begin{abstract}
We study standard and nonlocal nonlinear Schr\"{o}dinger (NLS) equations
obtained from the coupled NLS system of equations (Ablowitz-Kaup-Newell-Segur (AKNS)
equations) by using standard and nonlocal reductions respectively. By using the Hirota bilinear method we first find soliton solutions of the coupled NLS system of equations then using the reduction formulas we find the soliton solutions of the standard and nonlocal NLS equations. We give examples for particular values of the parameters and plot the function $|q(t,x)|^2$ for the standard and nonlocal NLS equations.

\end{abstract}

\section{Introduction}

When the Lax pairs are $sl(2,R)$ valued matrices (Ablowitz-Kaup-Newell-Segur (AKNS)
equations) and polynomials of the spectral parameter of degree two then the resulting equations are the following coupled nonlinear Schr\"{o}dinger equations \cite{abl0},
\begin{eqnarray}
a\,q_{t}&=& \frac{1}{2} q_{xx}-q^2\, r,\label{denk1}\\
a\,r_{t}&=&-\frac{1}{2} r_{xx}+r^2\, q, \label{denk2}
\end{eqnarray}
where $q(t,x)$ and $r(t,x)$ are complex dynamical variables, $a$ is a complex number in general. We call the above system of coupled equations as nonlinear Schr\"{o}dinger system (NLS system). The standard (local) reduction of this system
is obtained by letting
\begin{equation}
r(t,x)= k \bar{q}(t,x), \label{non0}
\end{equation}
where $k$ is a real constant and $\bar{q}$ is the complex conjugate of the function $q$. When this condition on the dynamical variables $q$ and $r$ is used in the system
of equations (\ref{denk1}) and (\ref{denk2}), they reduce to the following nonlinear Schr\"{o}dinger equation (NLS)
\begin{equation}
a\,q_{t}=\frac{1}{2} q_{xx}-k q^2\, \bar{q}, \label{denk3}
\end{equation}
provided that $\bar{a}=-a$. Recently, Ablowitz and Musslimani \cite{AbMu1}-\cite{AbMu3} found another integrable reduction. It is a nonlocal reduction of the NLS system (\ref{denk1}) and (\ref{denk2}) which is given by
\begin{equation}
r(t,x)=k \bar{q}(\varepsilon_{1} t, \varepsilon_{2} x), \label{non}
\end{equation}
where $(\varepsilon_{1})^2=(\varepsilon_{2})^2=1$. Under this
condition the NLS sytem (\ref{denk1}) and (\ref{denk2}) reduce to
\begin{equation}
a \,q_{t}(t,x)=\frac{1}{2} q_{xx}(t,x) -k q^2(t,x)\, \bar{q}(\varepsilon_{1} t, \varepsilon_{2} x), \label{denk4}
\end{equation}
provided that $\bar{a}=- \varepsilon_{1} a$. There is only one standard reduction where $(\varepsilon_{1}, \varepsilon_{2})=(1,1)$ but there are three different nonlocal reductions where $(\varepsilon_{1}, \varepsilon_{2})=\{(-1,1),(1,-1),(-1,-1)\}$. Hence for these values of $\varepsilon_{1}$ and  $\varepsilon_{2}$ and for different signs of $k$ (sign($k$)=$\pm 1$), we have six different nonlocal integrable NLS equations. They are respectively the time reflection symmetric (T-symmetric), the space reflection symmetric (S-symmetric), and the space-time reflection symmetric (ST-symmetric) nonlocal nonlinear Schr\"{o}dinger equations which are given by

\vspace{0.3cm}
\noindent
1. T-symmetric nonlinear Schr\"{o}dinger equation:
\begin{equation}\label{denk41}
a \,q_{t}(t,x)=\frac{1}{2} q_{xx}(t,x) - k q^2(t,x)\, \bar{q}(- t, x), ~~~\bar{a}=a.
\end{equation}

\vspace{0.3cm}
\noindent
2. S-symmetric nonlinear Schr\"{o}dinger equation:
\begin{equation}\label{denk42}
a \,q_{t}(t,x)=\frac{1}{2} q_{xx}(t,x)-k\, q^2(t,x)\, \bar{q}(t, -x),~~~\bar{a}=-a.
\end{equation}

\vspace{0.3cm}
\noindent
3. ST-symmetric nonlinear Schr\"{o}dinger equation:
\begin{equation}\label{denk43}
a \,q_{t}(t,x)=\frac{1}{2} q_{xx}(t,x) - k q^2(t,x)\, \bar{q}(-t, -x),~~~\bar{a}=a.
\end{equation}
Nonlocal NLS equations have the focusing and defocusing cases when the sign$(k)=-1$ and sign$(k)=1$ respectively.
All these equations are integrable. They possess Lax pairs and recursion operators.
In addition to the above equations (\ref{denk41})-(\ref{denk43}) we have also the equations for $q(-t,x)$, $q(t,-x)$, and $q(-t,-x)$ respectively. Since they are obtained from (\ref{denk41})-(\ref{denk43}) by $t \to -t$; $x \to -x$; and $(t \to -t, x \to -x)$ reflections respectively, we do not display them here.

Ablowitz and Musslimani have observed \cite{AbMu1} that one-soliton solutions of the nonlocal NLS equations blow up in a finite time. Existence of this singular behavior of one-soliton solutions of nonlocal NLS equations was also observed in Ref. \cite{gerd1}. Ablowitz and Musslimani have found many other nonlocal integrable equations such as nonlocal modified Korteweg-de Vries equation, nonlocal Davey-Stewartson equation, nonlocal sine-Gordon equation, and nonlocal $(2+1)$-dimensional three-wave interaction equations \cite{AbMu1}-\cite{AbMu3}. After the work of Ablowitz and Musslimani there is an increasing interest in obtaining the nonlocal reductions of systems of integrable equations and their properties \cite{fok}-\cite{SSL}.

The main purpose of this work is to search for possible integrable reductions of the NLS system (\ref{denk1}) and (\ref{denk2}) and investigate the applicability of the Hirota direct method to find the (soliton) solutions of the reduced nonlinear Schr\"{o}dinger equations.

By using the Hirota method we first find one- and two-soliton solutions of the NLS system of equations (\ref{denk1}) and (\ref{denk2}). We then investigate whether the system of equations (\ref{denk1}) and (\ref{denk2}) satisfy the Hirota integrability, i.e. existence of three-soliton solution \cite{H1}-\cite{H3}. We showed that the system possesses three-soliton solution. Then by using the reductions (\ref{non0}) and (\ref{non}) we obtain one-, two-, and also three-soliton solutions of the standard and nonlocal NLS equations, namely the equations (\ref{denk3}) and (\ref{denk41})-(\ref{denk43}) respectively. In this paper we give the general soliton solutions but we study only S-symmetric nonlocal NLS equations. We observe that all types of nonlocal NLS equations have singular and non-singular solutions depending on the values of the parameters in the solutions. In addition to the solitary wave solutions there are regular and singular localized solutions.  We give examples for certain values of the parameters and plot the function $|q(x,t)|^2$ for the S-symmetric case.

For the case S-symmetric nonlocal NLS equation (\ref{denk42}) we are at variance with Stalin et al.'s results \cite{SSL} (see Remark 2 and Remark 3 in Sections 4.1 and 4.2 respectively). They claim that they produce soliton solutions of the nonlocal NLS equation (S-symmetric) but it seems that they are solving the NLS system of equations (\ref{denk1}) and (\ref{denk2}) rather than solving nonlocal NLS equation (\ref{denk42}), because they ignore the constraint equations satisfied by the parameters of the one-soliton solutions.

The lay out of the paper is as follows. In Section 2 we apply Hirota method to the NLS system (\ref{denk1}) and (\ref{denk2}) and find one-, two-, and three-soliton solutions. In Section 3 we obtain soliton solutions of the standard NLS equation by using the standard reduction. In Section 4 we investigate soliton solutions of S-symmetric nonlocal NLS equation and give some examples for one-soliton, two-soliton, and three-soliton solutions and plot the function $|q(x,t)|^2$ for each example.

\section{Hirota Method for Coupled NLS System}

To find soliton solutions we use the Hirota method for (\ref{denk1}) and (\ref{denk2}). For this purpose we let
\begin{equation}\label{transformation}
\displaystyle q=\frac{F}{f}, \quad \quad r=\frac{G}{f}.
\end{equation}
The equation (\ref{denk1}) becomes
\begin{equation}
\displaystyle 2aF_tf^2-2aFf_tf-F_{xx}f^2+2F_xf_xf-2Ff_x^2+Ff_{xx}f+2GF^2=0,
\end{equation}
which is equivalent to
\begin{equation}
\displaystyle f(2aD_t-D_x^2)F\cdot f+F(D_x^2f\cdot f+2GF)=0.
\end{equation}
Similarly the equation (\ref{denk2}) becomes
\begin{equation}
\displaystyle 2aG_tf^2-2aGf_tf+G_{xx}f^2-2G_xf_xf+2Gf_x^2-Gf_{xx}f-2G^2F=0,
\end{equation}
which is equivalent to
\begin{equation}
\displaystyle f(2aD_t+D_x^2)G\cdot f-G(D_x^2f\cdot f+2GF)=0.
\end{equation}
Hence the Hirota bilinear form of the coupled NLS system (\ref{denk1}) and (\ref{denk2}) is
\begin{eqnarray}
P_1(D)\{F\cdot f\} \equiv (2aD_t-D_x^2+\alpha)\{F\cdot f\}&=&0\label{nFf}\\
P_2(D)\{G\cdot f\} \equiv (2aD_t+D_x^2-\alpha)\{G\cdot f\}&=&0\label{nGf}\\
P_3(D)\{f\cdot f\} \equiv (D_x^2-\alpha)\{f\cdot f\}&=&-2GF,\label{nff}
\end{eqnarray}
where $\alpha$ is an arbitrary constant.

\subsection{One-Soliton Solution of the NLS System}

To find one-soliton solution we use the following expansions for the functions $F$, $G$, and $f$,
\begin{equation}\label{expansion2}
F=\varepsilon F_1, \quad G=\varepsilon G_1, \quad f=1+\varepsilon^2 f_2,
\end{equation}
where
\begin{equation}\label{NLSF_1G_1}
F_1=e^{\theta_1}, \quad G_1=e^{\theta_2}, \quad \theta_i=k_ix+\omega_i t+\delta_i,\, i= 1, 2.
\end{equation}
When we substitute (\ref{expansion2}) into the equations (\ref{nFf})-(\ref{nff}), we obtain the
coefficients of $\varepsilon$ as
\begin{eqnarray}
&&P_1(D)\{F_1\cdot 1\}=2aF_{1,t}-F_{1,xx}+\alpha F_1=0,\\
&&P_2(D)\{G_1\cdot 1\}=2aG_{1,t}+G_{1,xx}-\alpha G_1=0,
\end{eqnarray}
yielding the dispersion relations
\begin{equation}\label{dispersionNLSone}
\omega_1=\frac{(k_1^2-\alpha)}{2a},\quad \omega_2=\frac{(\alpha-k_2^2)}{2a}.
\end{equation}
From the coefficient of $\varepsilon^2$
\begin{equation}
f_{2,xx}-\alpha f_2=-G_1F_1,
\end{equation}
we obtain the function $f_2$ as
\begin{equation}\label{NLSf_2}
\displaystyle f_2=\frac{ e^{(k_1+k_2)x+(\omega_1+\omega_2)t+\delta_1+\delta_2} }{\alpha-(k_1+k_2)^2}.
\end{equation}
The coefficients of $\varepsilon^3$ vanish due to the dispersion relations and (\ref{NLSf_2}). From the coefficient of
$\varepsilon^4$
\begin{equation}
(D_x^2-\alpha)\{f_2\cdot f_2\}=2(f_2f_{2,xx}-f_{2,x}^2)-\alpha f_2^2=0,
\end{equation}
by using the function $f_2$ given in (\ref{NLSf_2}), we get that $\alpha=0$. In the rest of the paper we will take $\alpha=0$. Let us also take $\varepsilon=1$. Hence a pair of solutions of the NLS system (\ref{denk1}) and (\ref{denk2})
is given by $(q(t,x),r(t,x))$ where
\begin{equation}\label{NLSsystemonesol}
\displaystyle q(t,x)=\frac{e^{\theta_1}}{1+Ae^{\theta_1+\theta_2}}, \quad \quad r(t,x)=\frac{e^{\theta_2}}{1+Ae^{\theta_1+\theta_2}},
\end{equation}
with $\theta_i=k_ix+\omega_it+\delta_i$, $i=1,2$, $\displaystyle \omega_1=\frac{k_1^2}{2a}$, $\displaystyle \omega_2=-\frac{k_2^2}{2a}$, and $\displaystyle A=-\frac{1}{(k_1+k_2)^2}$. Here $k_{1}$, $k_{2}$, $\delta_{1},$ and $\delta_{2}$ are arbitrary complex numbers.

\subsection{Two-Soliton Solution of the NLS System}

For two-soliton solution, we take
\begin{equation}\label{expansionNLS2}
f=1+\varepsilon^2 f_2+\varepsilon^4 f_4,\quad G=\varepsilon G_1+\varepsilon^3 G_3,\quad F=\varepsilon F_1+\varepsilon^3 F_3,
\end{equation}
where
\begin{equation}\label{funcNLS2}
F_1=e^{\theta_1}+e^{\theta_2}, \quad G_1=e^{\eta_1}+e^{\eta_2},
\end{equation}
with $\theta_i=k_i x+\omega_i t+\delta_i$, $\eta_i=\ell_ix+m_it+\alpha_i$ for $i=1, 2$. When we insert the above expansions into (\ref{nFf})-(\ref{nff})
we obtain the coefficients of $\varepsilon$ as
\begin{align}
&P_1(D)\{F_1\cdot 1\}=2aF_{1,t}-F_{1,xx}=0,\\
&P_2(D)\{G_1\cdot 1\}=2aG_{1,t}+G_{1,xx}=0.
\end{align}
Here we get the dispersion relations
\begin{equation}\label{dispersion}
\omega_i=\frac{k_i^2}{2a},\quad  m_i=-\frac{\ell_i^2}{2a},\quad i=1, 2.
\end{equation}
The coefficient of $\varepsilon^2$ gives
\begin{equation}
f_{2,xx}=-G_1F_1
\end{equation}
yielding the function $f_2$
\begin{equation}\displaystyle
f_2= e^{\theta_1+\eta_1+\alpha_{11}}+e^{\theta_1+\eta_2+\alpha_{12}}+e^{\theta_2+\eta_1+\alpha_{21}}+e^{\theta_2+\eta_2+\alpha_{22}}=\sum_{1\leq i,j\leq 2}e^{\theta_i+\eta_j+\alpha_{ij}},
\end{equation}
where
\begin{equation}
\displaystyle e^{\alpha_{ij}}=-\frac{1}{(k_i+\ell_j)^2},\, 1\leq i,j\leq 2.
\end{equation}
From the coefficients of $\varepsilon^3$ we get
\begin{align}
&2a(F_{1,t}f_2-F_1f_{2,t})-F_{1,xx}f_2+2F_{1,x}f_{2,x}-F_1f_{2,xx}+2aF_{3,t}-F_{3,xx}=0,\\
&2a(G_{1,t}f_2-G_1f_{2,t})+G_{1,xx}f_2-2G_{1,x}f_{2,x}+G_1f_{2,xx}+2aG_{3,t}+G_{3,xx}=0.
\end{align}
These equations give the functions $F_3$ and $G_3$ as
\begin{equation}
F_3=A_1e^{\theta_1+\theta_2+\eta_1}+A_2e^{\theta_1+\theta_2+\eta_2},\quad G_3=B_1e^{\theta_1+\eta_1+\eta_2}+B_2e^{\theta_2+\eta_1+\eta_2},
\end{equation}
where
\begin{equation}\label{A_iB_i}
\displaystyle A_i=-\frac{(k_1-k_2)^2}{(k_1+\ell_i)^2(k_2+\ell_i)^2}, \quad B_i=-\frac{(\ell_1-\ell_2)^2}{(\ell_1+k_i)^2(\ell_2+k_i)^2},\quad i=1, 2.
\end{equation}
The coefficient of $\varepsilon^4$ gives
\begin{equation}
f_{4,xx}+(f_2f_{2,xx}-f_{2,x}^2)+G_1F_3+G_3F_1=0,
\end{equation}
yielding the function $f_4$ as
\begin{equation}
f_4=Me^{\theta_1+\theta_2+\eta_1+\eta_2},
\end{equation}
where
\begin{equation}\label{M}
\displaystyle M=\frac{(k_1-k_2)^2(l_1-l_2)^2}{(k_1+l_1)^2(k_1+l_2)^2(k_2+l_1)^2(k_2+l_2)^2}.
\end{equation}
The coefficients of $\varepsilon^5$;
\begin{align*}
&2a(F_{3,t}f_2-F_3f_{2,t})-F_{3,xx}f_2+2F_{3,x}f_{2,x}-F_3f_{2,xx}+2a(F_{1,t}f_4-F_1f_{4,t})\\
&-F_{1,xx}f_4+2F_{1,x}f_{4,x}-F_1f_{4,xx}=0,\\
&2a(G_{3,t}f_2-G_3f_{2,t})+G_{3,xx}f_2-2G_{3,x}f_{2,x}+G_3f_{2,xx}+2a(G_{1,t}f_4-G_1f_{4,t})\\
&+G_{1,xx}f_4-2G_{1,x}f_{4,x}+G_1f_{4,xx}=0,\\
\end{align*}
the coefficient of $\varepsilon^6$;
\begin{equation*}
f_{2,xx}f_4-2f_{2,x}f_{4,x}+f_2f_{4,xx}+G_3F_3=0,
\end{equation*}
the coefficients of $\varepsilon^7$;
\begin{eqnarray*}
&&2a(F_{3,t}f_4-F_3f_{4,t})-F_{3,xx}f_4+2F_{3,x}f_{4,x}-F_3f_{4,xx}=0,\\
&&2a(G_{3,t}f_4-G_3f_{4,t})+G_{3,xx}f_4-2G_{3,x}f_{4,x}+G_3f_{4,xx}=0,
\end{eqnarray*}
and the coefficient of $\varepsilon^8$;
\begin{equation*}
f_4f_{4,xx}-f_{4,x}^2=0,
\end{equation*}
vanish directly due to the functions $F_1, G_1$, and $F_3, G_3, f_2, f_4$ that are previously found. If we take $\varepsilon=1$ then two-soliton solution of the
NLS system (\ref{denk1}) and (\ref{denk2}) is given with the pair $(q(t,x),r(t,x))$ where
\begin{align}
\displaystyle& q(t,x)=\frac{e^{\theta_1}+e^{\theta_2}+A_1e^{\theta_1+\theta_2+\eta_1}+A_2e^{\theta_1+\theta_2+\eta_2}}
{1+e^{\theta_1+\eta_1+\alpha_{11}}+e^{\theta_1+\eta_2+\alpha_{12}}+e^{\theta_2+\eta_1+\alpha_{21}}+e^{\theta_2+\eta_2+\alpha_{22}}
+Me^{\theta_1+\theta_2+\eta_1+\eta_2}},\label{NLS2solq(t,x)}\\
&r(t,x)=\frac{e^{\eta_1}+e^{\eta_2}+B_1e^{\theta_1+\eta_1+\eta_2}+B_2e^{\theta_2+\eta_1+\eta_2}}
{1+e^{\theta_1+\eta_1+\alpha_{11}}+e^{\theta_1+\eta_2+\alpha_{12}}+e^{\theta_2+\eta_1+\alpha_{21}}+e^{\theta_2+\eta_2+\alpha_{22}}
+Me^{\theta_1+\theta_2+\eta_1+\eta_2}},\label{NLS2solr(t,x)}
\end{align}
with $\displaystyle \theta_i=k_ix+\frac{k_i^2}{2a}t+\delta_i$, $\displaystyle \eta_i=\ell_ix-\frac{\ell_i^2}{2a}t+\alpha_i$ for $i=1, 2$. Here $k_{i}$, $\ell_{i}, \delta_{i}$, and $\alpha_{i}$, $i=1, 2$ are arbitrary complex numbers.
\subsection{Three-Soliton Solution of the NLS System}

Hirota integrability is defined as the existence of three-soliton solutions. For this purpose we
find three-soliton solutions of the NLS system (\ref{denk1}) and (\ref{denk2}) and all of its reductions.

For three-soliton solution, we take
\begin{equation}\label{expansionNLS3}
f=1+\varepsilon^2 f_2+\varepsilon^4 f_4+\varepsilon^6 f_6,\quad G=\varepsilon G_1+\varepsilon^3 G_3+\varepsilon^5 G_5,\quad F=\varepsilon F_1+\varepsilon^3 F_3+\varepsilon^5 F_5,
\end{equation}
and
\begin{equation}\label{funcNLS2}
F_1=e^{\theta_1}+e^{\theta_2}+e^{\theta_3}, \quad G_1=e^{\eta_1}+e^{\eta_2}+e^{\eta_3},
\end{equation}
where $\theta_i=k_i x+\omega_i t+\delta_i$, $\eta_i=\ell_ix+m_it+\alpha_i$ for $i=1, 2, 3$. We insert the expansions to the Hirota bilinear form of NLS system
(\ref{nFf})-(\ref{nff}) and obtain the coefficients of $\varepsilon^n$, $1\leq n \leq 12$ as
\begin{align}
\varepsilon:&\, 2aF_{1,t}-F_{1,xx}=0,\label{dispF}\\
&\, 2aG_{1,t}+G_{1,xx}=0,\label{dispG}\\
\varepsilon^2:&\, f_{2,xx}+G_1F_1=0,\label{NLS3f_2}\\
\varepsilon^3:&\, 2a(F_{1,t}f_2-F_1f_{2,t})-F_{1,xx}f_2+2F_{1,x}f_{2,x}-F_1f_{2,xx}+2aF_{3,t}-F_{3,xx}=0,\\
&\, 2a(G_{1,t}f_2-G_1f_{2,t})+G_{1,xx}f_2-2G_{1,x}f_{2,x}+G_1f_{2,xx}+2aG_{3,t}+G_{3,xx}=0,\\
\varepsilon^4:&\, f_{4,xx}+f_2f_{2,xx}-f_{2,x}^2+G_1F_3+G_3F_1=0,\label{NLSf_4}
\end{align}
\begin{align}
\varepsilon^5:&\, 2a(F_{3,t}f_2-F_3f_{2,t})-F_{3,xx}f_2+2F_{3,x}f_{2,x}-F_3f_{2,xx}+2a(F_{1,t}f_4-F_1f_{4,t})\nonumber\\
&-F_{1,xx}f_4+2F_{1,x}f_{4,x}-F_1f_{4,xx}+2aF_{5,t}-F_{5,xx}=0,\label{NLSF5}\\
&\, 2a(G_{3,t}f_2-G_3f_{2,t})+G_{3,xx}f_2-2G_{3,x}f_{2,x}+G_3f_{2,xx}+2a(G_{1,t}f_4-G_1f_{4,t})\nonumber\\
&+G_{1,xx}f_4-2G_{1,x}f_{4,x}+G_1f_{4,xx}+2aG_{5,t}+G_{5,xx}=0,\label{NLSG5}\\
\varepsilon^6:&\, f_{2,xx}f_4-2f_{2,x}f_{4,x}+f_2f_{4,xx}+f_{6,xx}+G_5F_1+G_1F_5+G_3F_3=0,\label{NLSf6}\\
\varepsilon^7:&\, 2a(F_{3,t}f_4-F_3f_{4,t})-F_{3,xx}f_4+2F_{3,x}f_{4,x}-F_3f_{4,xx}+2a(F_{1,t}f_6-F_1f_{6,t})\nonumber\\
&-F_{1,xx}f_6+2F_{1,x}f_{6,x}-F_1f_{6,xx}+2a(F_{5,t}f_2-F_5f_{2,t})-F_{5,xx}f_2\nonumber\\
&+2F_{5,x}f_{2,x}-F_5f_{2,xx}=0,\label{rest1}\\
&\, 2a(G_{3,t}f_4-G_3f_{4,t})+G_{3,xx}f_4-2G_{3,x}f_{4,x}+G_3f_{4,xx}+2a(G_{1,t}f_6-G_1f_{6,t})\nonumber\\
&+G_{1,xx}f_6-2G_{1,x}f_{6,x}+G_1f_{6,xx}+2a(G_{5,t}f_2-G_5f_{2,t})+G_{5,xx}f_2\nonumber\\
&-2G_{5,x}f_{2,x}+G_5f_{2,xx}=0,\label{rest2}\\
\varepsilon^8:&\, f_{2,xx}f_6-2f_{2,x}f_{6,x}+f_2f_{6,xx}+f_4f_{4,xx}-f_{4,x}^2+G_3F_5+G_5F_3=0,\label{rest3}\\
\varepsilon^9:&\, 2a(F_{3,t}f_6-F_3f_{6,t})-F_{3,xx}f_6+2F_{3,x}f_{6,x}-F_3f_{6,xx}+2a(F_{5,t}f_4-F_5f_{4,t})\nonumber\\
&-F_{5,xx}f_4+2F_{5,x}f_{4,x}-F_5f_{4,xx}=0,\label{rest4}\\
&\, 2a(G_{3,t}f_6-G_3f_{6,t})+G_{3,xx}f_6-2G_{3,x}f_{6,x}+G_3f_{6,xx}+2a(G_{5,t}f_4-G_5f_{4,t})\nonumber\\
&+G_{5,xx}f_4-2G_{5,x}f_{4,x}+G_5f_{4,xx}=0,\label{rest5}\\
\varepsilon^{10}:&\, f_{4,xx}f_6-2f_{4,x}f_{6,x}+f_4f_{6,xx}+G_5F_5=0,\label{rest6}\\
\varepsilon^{11}:&\, 2a(F_{5,t}f_6-F_5f_{6,t})-F_{5,xx}f_6+2F_{5,x}f_{6,x}-F_5f_{6,xx}=0,\label{rest7}\\
&\,  2a(G_{5,t}f_6-G_5f_{6,t})+G_{5,xx}f_6-2G_{5,x}f_{6,x}+G_5f_{6,xx}=0,\label{rest8}\\
\varepsilon^{12}:&\, f_6f_{6,xx}-f_{6,x}^2=0.\label{rest9}
\end{align}
From the equalities (\ref{dispF}) and (\ref{dispG}) we obtain the dispersion relations
\begin{equation}\label{dispersionNLS3}
\omega_i=\frac{k_i^2}{2a},\quad  m_i=-\frac{\ell_i^2}{2a},\quad i=1, 2, 3.
\end{equation}
Equation (\ref{NLS3f_2}) gives the function $f_2$ as
\begin{equation}\displaystyle
f_2=\sum_{1\leq i, j\leq 3} e^{\theta_i+\eta_j+\alpha_{ij}}, \quad e^{\alpha_{ij}}=-\frac{1}{(k_i+\ell_j)^2},\quad 1\leq i, j\leq 3.
\end{equation}
From the coefficients of $\varepsilon^3$, we obtain the functions $F_3$ and $G_3$
\begin{align}\displaystyle
&F_3=\sum_{\substack{1\leq i,j,s\leq 3 \\ i<j}} A_{ijs}e^{\theta_i+\theta_j+\eta_s}, \quad A_{ijs}=-\frac{(k_i-k_j)^2}{(k_i+\ell_s)^2(k_j+\ell_s)^2},\quad 1\leq i,j,s\leq 3, i<j,\\
&G_3=\sum_{\substack{1\leq i,j,s\leq 3 \\ i<j}} B_{ijs}e^{\eta_i+\eta_j+\theta_s}, \quad B_{ijs}=-\frac{(\ell_i-\ell_j)^2}{(\ell_i+k_s)^2(\ell_j+k_s)^2},\quad 1\leq i,j,s\leq 3, i<j.
\end{align}
The equation (\ref{NLSf_4}) yields the function $f_4$ as
\begin{equation}
f_4=\sum_{\substack{1\leq i<j\leq 3 \\ 1\leq p<r \leq3}} M_{ijpr}e^{\theta_i+\theta_j+\eta_p+\eta_r},
\end{equation}
where
\begin{equation}\displaystyle
M_{ijpr}=\frac{(k_i-k_j)^2(l_p-l_r)^2}{(k_i+l_p)^2(k_i+l_r)^2(k_j+l_p)^2(k_j+l_r)^2},
\end{equation}
for $1\leq i< j\leq 3$, $1\leq p< r\leq 3$. From the coefficients of $\varepsilon^5$ we obtain the functions $F_5$ and $G_5$,
\begin{align}
&F_5=V_{12}e^{\theta_1+\theta_2+\theta_3+\eta_1+\eta_2}+V_{13}e^{\theta_1+\theta_2+\theta_3+\eta_1+\eta_3}+V_{23}e^{\theta_1+\theta_2+\theta_3+\eta_2+\eta_3},\\
&G_5=W_{12}e^{\theta_1+\theta_2+\eta_1+\eta_2+\eta_3}+W_{13}e^{\theta_1+\theta_2+\eta_1+\eta_2+\eta_3}+W_{23}e^{\theta_2+\theta_3+\eta_1+\eta_2+\eta_3},
\end{align}
where
\begin{eqnarray}\displaystyle
&&V_{ij}=\frac{S_{ij}}{(k_1+k_2+k_3+\ell_i+\ell_j)^2-2a(\omega_1+\omega_2+\omega_3+m_i+m_j)},\\
 &&W_{ij}=-\frac{Q_{ij}}{(k_i+k_j+\ell_1+\ell_2+\ell_3)^2+2a(\omega_i+\omega_j+m_1+m_2+m_3)},
\end{eqnarray}
for $1\leq i<j\leq 3$. Here $S_{ij}$ and $Q_{ij}$ are given in Appendix of Ref. \cite{GurPek}.
The equation (\ref{NLSf6}) gives the function $f_6$
\begin{equation}\displaystyle
f_6=He^{\theta_1+\theta_2+\theta_3+\eta_1+\eta_2+\eta_3},
\end{equation}
where the coefficient $H$ is also represented in Appendix of Ref. \cite{GurPek}. The rest of the equations (\ref{rest1})-(\ref{rest9})
are satisfied directly. Let us also take $\varepsilon=1$. Hence three-soliton solution of the coupled NLS system (\ref{denk1}) and (\ref{denk2}) is given with the pair $(q(t,x),r(t,x))$ where
\begin{align}\displaystyle
q(t,x)&=\frac{e^{\theta_1}+e^{\theta_2}+e^{\theta_3}+\sum_{\substack{1\leq i,j,s\leq 3 \\ i<j}} A_{ijs}e^{\theta_i+\theta_j+\eta_s}
+\sum_{\substack{1\leq i,j\leq 3 \\ i<j}}V_{ij}e^{\theta_1+\theta_2+\theta_3+\eta_i+\eta_j}}{1+\sum_{1\leq i,j\leq3}e^{\theta_i+\eta_j+\alpha_{ij}}+\sum_{\substack{1\leq i<j\leq 3 \\ 1\leq p<r \leq3}} M_{ijpr}e^{\theta_i+\theta_j+\eta_p+\eta_r}+He^{\theta_1+\theta_2+\theta_3+\eta_1+\eta_2+\eta_3}}, \\
r(t,x)&=\frac{e^{\eta_1}+e^{\eta_2}+e^{\eta_3}+\sum_{\substack{1\leq i,j,s\leq 3 \\ i<j}} B_{ijs}e^{\eta_i+\eta_j+\theta_s}+
\sum_{\substack{1\leq i,j\leq 3 \\ i<j}}W_{ij}e^{\theta_i+\theta_j+\eta_1+\eta_2+\eta_3}}{1+\sum_{1\leq i,j\leq3}e^{\theta_i+\eta_j+\alpha_{ij}}+\sum_{\substack{1\leq i<j\leq 3 \\ 1\leq p<r \leq3}} M_{ijpr}e^{\theta_i+\theta_j+\eta_p+\eta_r}+He^{\theta_1+\theta_2+\theta_3+\eta_1+\eta_2+\eta_3}}.\label{NLS3solr(t,x)}
\end{align}
\vspace{0.3cm}

\noindent \textbf{Remark 1.} Notice that the authors of Ref. \cite{SSL} used another form of Hirota perturbation expansion for one-soliton solution;
\begin{equation}\displaystyle
q(t,x)=\frac{g(t,x)}{f(t,x)},
\end{equation}
where
\begin{equation}
g(t,x)=\varepsilon g_1+\varepsilon^3g_3, \quad f(t,x)=1+\varepsilon^2f_2+\varepsilon^4f_4,
\end{equation}
different than the form (\ref{expansion2}) that we use. The solution found in Ref. \cite{SSL},
\begin{equation}\label{SSLclosedonesol}\displaystyle
  q(t,x)=\frac{ \alpha_1e^{\bar{\xi}_1}+e^{\xi_1+2\bar{\xi}_1+\delta_{11}}}{1+e^{\xi_1+\bar{\xi}_1+\delta_1}+e^{2(\xi_1+\bar{\xi}_1)+R}},
\end{equation}
the numerator and denominator are factorizable and it reduces to our solution (\ref{NLSsystemonesol})
\begin{equation}\label{SSLclosedonesol}\displaystyle
  q(t,x)= \frac{\alpha_1 e^{\bar{\xi}_1}\,(1+\frac{1}{\alpha_{1}}\, e^{\xi_1+\bar{\xi}_1+\delta_{11}})}{(1+e^{\xi_1+\bar{\xi}_1+\Delta}\,)(1+\frac{1}{\alpha_{1}}\, e^{\xi_1+\bar{\xi}_1+\delta_{11}})}=\frac{\alpha_1e^{\bar{\xi}_1}}{1+e^{\xi_1+\bar{\xi}_1+\Delta}}.
\end{equation}
For two-soliton solution, the following form of Hirota perturbation expansion
\begin{equation}
g(t,x)=\sum_{n=0}^3\varepsilon^{2n+1}g_{2n+1}, \quad f(t,x)=1+\sum_{n=1}^{4}\varepsilon^{2n}f_{2n},
\end{equation}
is used in Ref. \cite{SSL}. Our two-soliton solutions (\ref{NLS2solq(t,x)}) and (\ref{NLS2solr(t,x)}) are much simpler and shorter than the one given in \cite{SSL}. Similar to one-soliton solution, one expects that the two-soliton solution given in Ref. \cite{SSL} is equivalent to the solutions (\ref{NLS2solq(t,x)}) and (\ref{NLS2solr(t,x)}).

\section{Standard Reduction of the NLS System}

Here we consider the standard reduction (\ref{non0}) and obtain soliton solutions of
the reduced equation (\ref{denk3}) with the condition
\begin{equation}\label{standardredNLScond}
 \bar{a}=-a
\end{equation}
satisfied.

\subsection{One-Soliton Solution for the Standard NLS Equation}
We first obtain the conditions on the parameters of one-soliton solution of the NLS system to satisfy the equality (\ref{non0}) i.e.,
\begin{equation}
\displaystyle \frac{e^{k_2x-\frac{k_2^2}{2a}t+\delta_2}}{1+Ae^{(k_1+k_2)x+\frac{(k_1^2-k_2^2)}{2a}t+\delta_1+\delta_2  }}
=k\frac{e^{\bar{k}_1x+\frac{\bar{k}_1^2}{2\bar{a}}t+\bar{\delta}_1}}
{1+\bar{A}e^{(\bar{k}_1+\bar{k}_2)x+\frac{(\bar{k}_1^2-\bar{k}_2^2)}{2\bar{a}}t+\bar{\delta}_1+\bar{\delta}_2}}.
\end{equation}
Hence one of the set of the constraints that the parameters must satisfy is the following:
\begin{align}\label{stanoneconds}
\displaystyle &i)\, k_2=\bar{k}_1,\quad  ii)\,-\frac{k_2^2}{2a}=\frac{\bar{k}_1^2}{2\bar{a}},\quad iii)\, e^{\delta_2}=ke^{\bar{\delta}_1},\quad
iv)\, A=\bar{A},\nonumber\\
 &v)\,(k_1+k_2)=(\bar{k}_1+\bar{k}_2),\quad
 vi)\,\frac{(k_1^2-k_2^2)}{2a}=\frac{(\bar{k}_1^2-\bar{k}_2^2)}{2\bar{a}},\quad vii)\, e^{\delta_1+\delta_2}=e^{\bar{\delta}_1+\bar{\delta}_2}.
\end{align}
Consider the condition $ii)$. We have
\begin{equation}\displaystyle
-\frac{k_2^2}{2a}=-\frac{\bar{k}_1}{-2\bar{a}}=\frac{\bar{k}_1^2}{2\bar{a}},
\end{equation}
by (\ref{standardredNLScond}) and the condition $i)$. Similarly, the conditions $iv)-vi)$ are also satisfied directly by
(\ref{standardredNLScond}) and $i)$.
Now consider the relation $e^{\delta_2}=ke^{\bar{\delta}_1}$ or $e^{\bar{\delta}_2}=ke^{\delta_1}$ given in $iii)$ of (\ref{stanoneconds}). Note that
since $k$ is a real constant we have $\bar{k}=k$. Consequently, we have
\begin{equation*}
e^{\delta_1+\delta_2}=ke^{\delta_1}e^{\bar{\delta}_1}\quad \mathrm{and} \quad e^{\bar{\delta}_1+\bar{\delta}_2}=ke^{\bar{\delta}_1}e^{\delta_1},
\end{equation*}
yielding the equality $e^{\delta_1+\delta_2}=e^{\bar{\delta}_1+\bar{\delta}_2}$.

\noindent Therefore the parameters of
one-soliton solution of the equation (\ref{denk3}) must have the following properties:
\begin{equation}
1)\,\bar{a}=-a,\quad 2)\,k_2=\bar{k}_1,\quad 3)\, e^{\delta_2}=ke^{\bar{\delta}_1}.
\end{equation}

\vspace{0.3cm}
\noindent
{\bf Example 1.}\, Let us illustrate a particular example of one-soliton solution of (\ref{denk3}). For $(k_1,k_2, e^{\delta_1}, e^{\delta_2}, k, a)= (1+i,1-i,i,i,-1,\frac{i}{2})$,
one-soliton solution becomes
\begin{equation}\displaystyle
q(t,x)=\frac{ie^{(1+i)x+2t}}{1+\frac{1}{4}e^{2x+4t}}.
\end{equation}
\noindent To sketch the graph of this solution in real plane we will consider $q(t,x)\bar{q}(t,x)=|q(t,x)|^2$,
\begin{equation}\label{stanex1}
|q(t,x)|^2=\frac{16e^{2x+4t}}{(4+e^{2x+4t})^2}.
\end{equation}
The graph of (\ref{stanex1}) is given in Figure 1.
 \begin{center}
\begin{figure}[h!]
\centering
\begin{minipage}[t]{1\linewidth}
\centering
\includegraphics[angle=0,scale=.24]{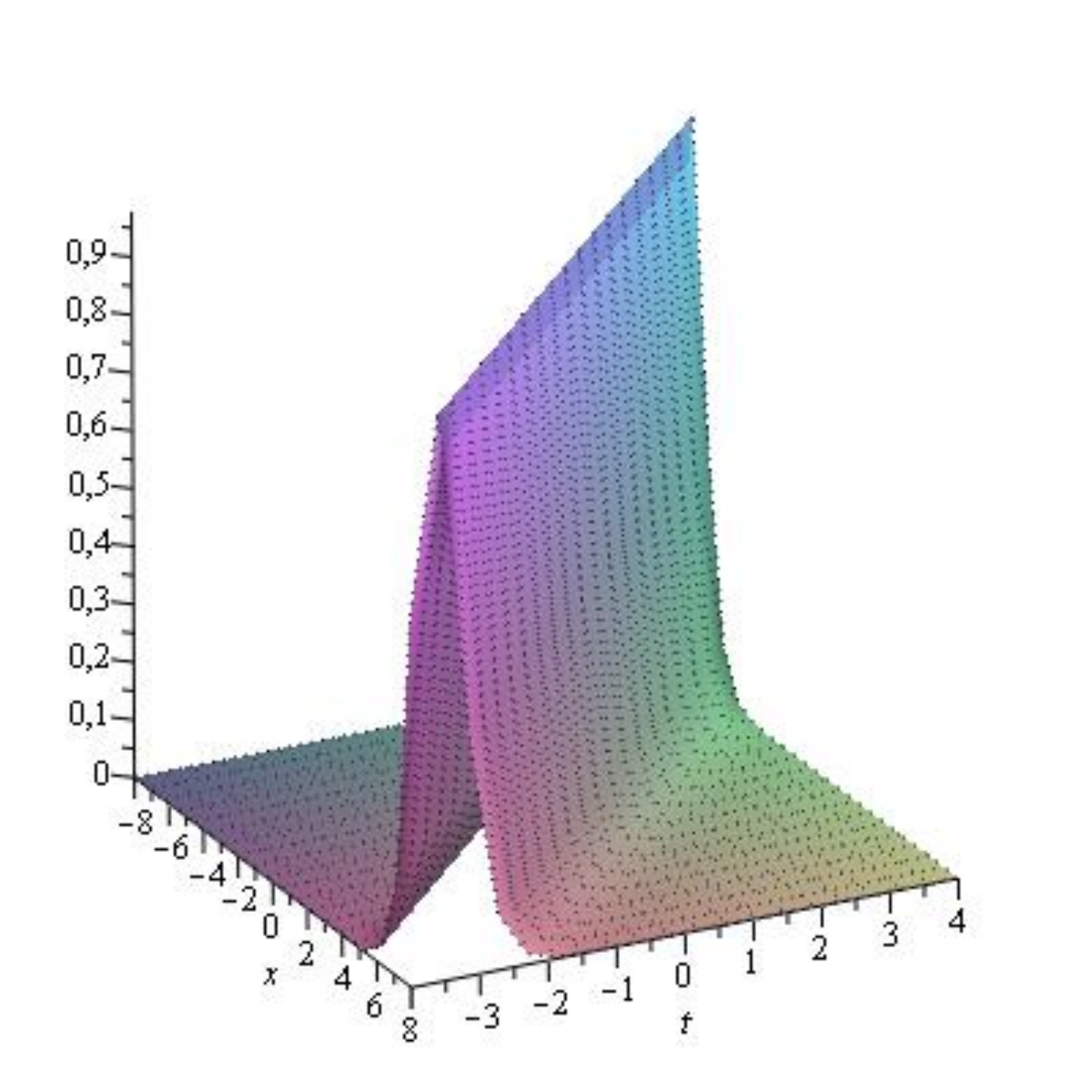}
\caption{One-soliton solution for (\ref{stanex1}).}
\end{minipage}
\end{figure}
\end{center}
\subsection{Two-Soliton Solution for the Standard NLS Equation}
Similar to one-soliton solution case, we obtain the conditions on the parameters of two-soliton solution given by (\ref{NLS2solq(t,x)}) and (\ref{NLS2solr(t,x)}) of the NLS system to satisfy the equality (\ref{non0});
\begin{equation}
1)\, \bar{a}=-a,\quad 2)\, \ell_i=\bar{k}_i, \,\, i=1, 2,\quad 3)\, e^{\alpha_i}=ke^{\bar{\delta}_i},\,\, i=1, 2.
\end{equation}

\vspace{0.3cm}
\noindent
{\bf Example 2.} \, Consider the following parameters: $(k_1,\ell_1,k_2,\ell_2)=(1+i,1-i,2+2i,2-2i)$ with $(e^{\alpha_j},e^{\delta_j}, k, a)=(-1+i,1+i,-1,i)$ for $j=1, 2$. In this case two-soliton solution
is
\begin{equation}\label{stantwosol1}\displaystyle
q(t,x)=\frac{Y_1}{Y_2},
\end{equation}
where
\begin{equation*}\displaystyle
Y_1=(1+i)e^{(1+i)x+t}+(1+i)e^{(2+2i)x+4t}+\Big(-\frac{1}{50}+\frac{7}{50}i\Big)e^{(4+2i)x+6t}
+\Big(-\frac{7}{200}+\frac{1}{200}i\Big)e^{(5+i)x+9t},
\end{equation*}
and
\begin{equation*}\displaystyle
Y_2=1+\frac{1}{2}e^{2x+2t}+\Big(\frac{4}{25}+\frac{3}{25}i\Big)e^{(3-i)x+5t}+\Big(\frac{4}{25}-\frac{3}{25}i\Big)e^{(3+i)x+5t}
+\frac{1}{8}e^{4x+8t}+\frac{1}{400}e^{6x+10t}.
\end{equation*}
The graph of the function $|q(t,x)|^2$ corresponding to the solution (\ref{stantwosol1}) is
given in Figure 2.a.

\vspace{0.3cm}
\noindent
{\bf Example 3.} \, In this example we just give the graphs of two-soliton solutions defined by the function $|q(t,x)|^2$ corresponding to
$ (k_1, \ell_1, k_2, \ell_2)=\Big(-\frac{1}{2}-\frac{2}{5}i,-\frac{1}{2}+\frac{2}{5}i, -\frac{13}{25}+\frac{2}{5}i,-\frac{13}{25}-\frac{2}{5}i\Big)$ and $ (k_1, \ell_1, k_2, \ell_2)=\Big(-\frac{1}{2}-\frac{2}{5}i,-\frac{1}{2}+\frac{2}{5}i, \frac{13}{25}-\frac{2}{5}i,\frac{13}{25}+\frac{2}{5}i\Big)$ with $(e^{\alpha_j},e^{\delta_j}, k, a)=(-1+i,1+i,-1,i)$ for $j=1, 2$ in Figures 2.b and 2.c, respectively.
\begin{figure}[h!]
\centering     
\subfigure[]{\label{fig:2a}\includegraphics[width=34mm]{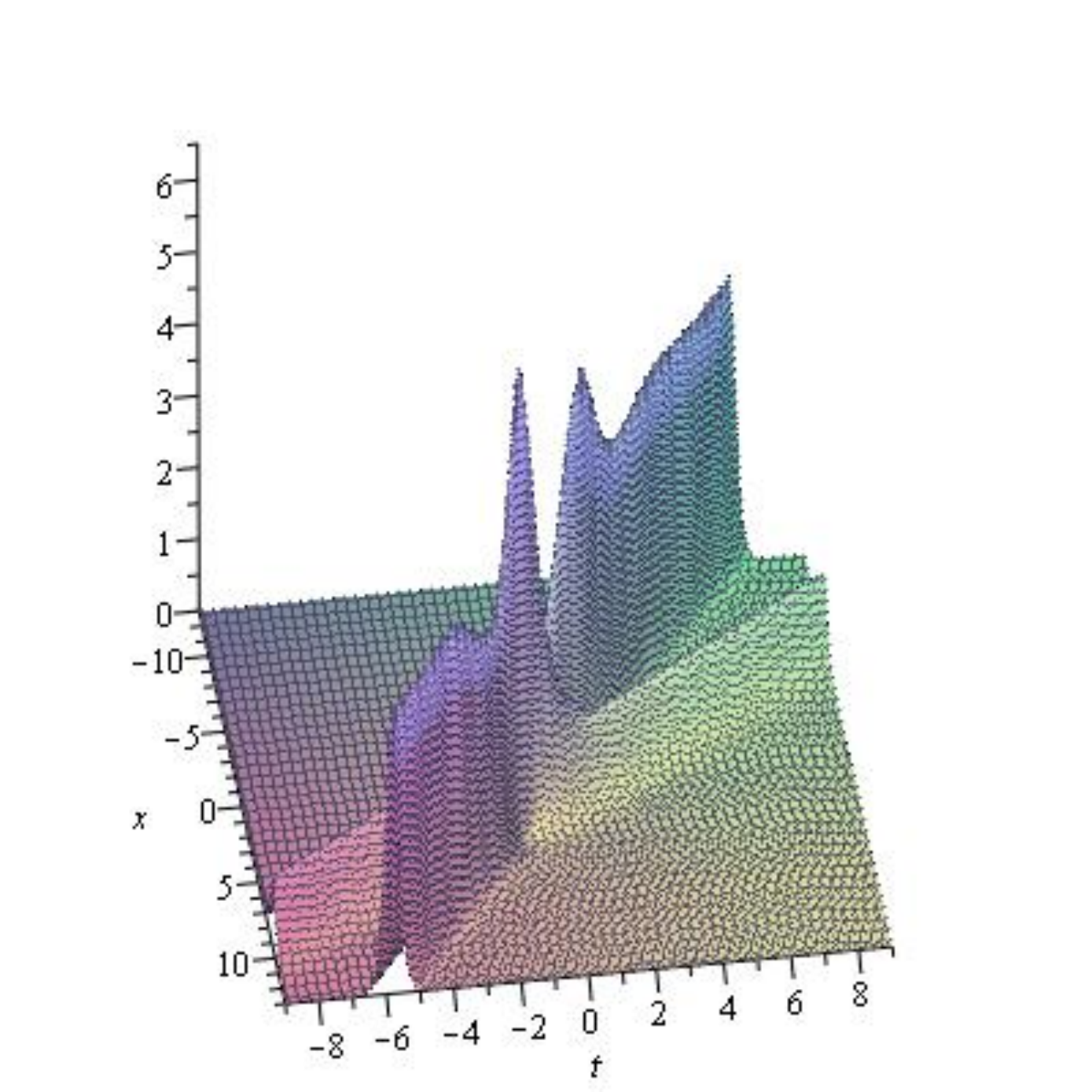}}\hfill
\subfigure[]{\label{fig:2b}\includegraphics[width=34mm]{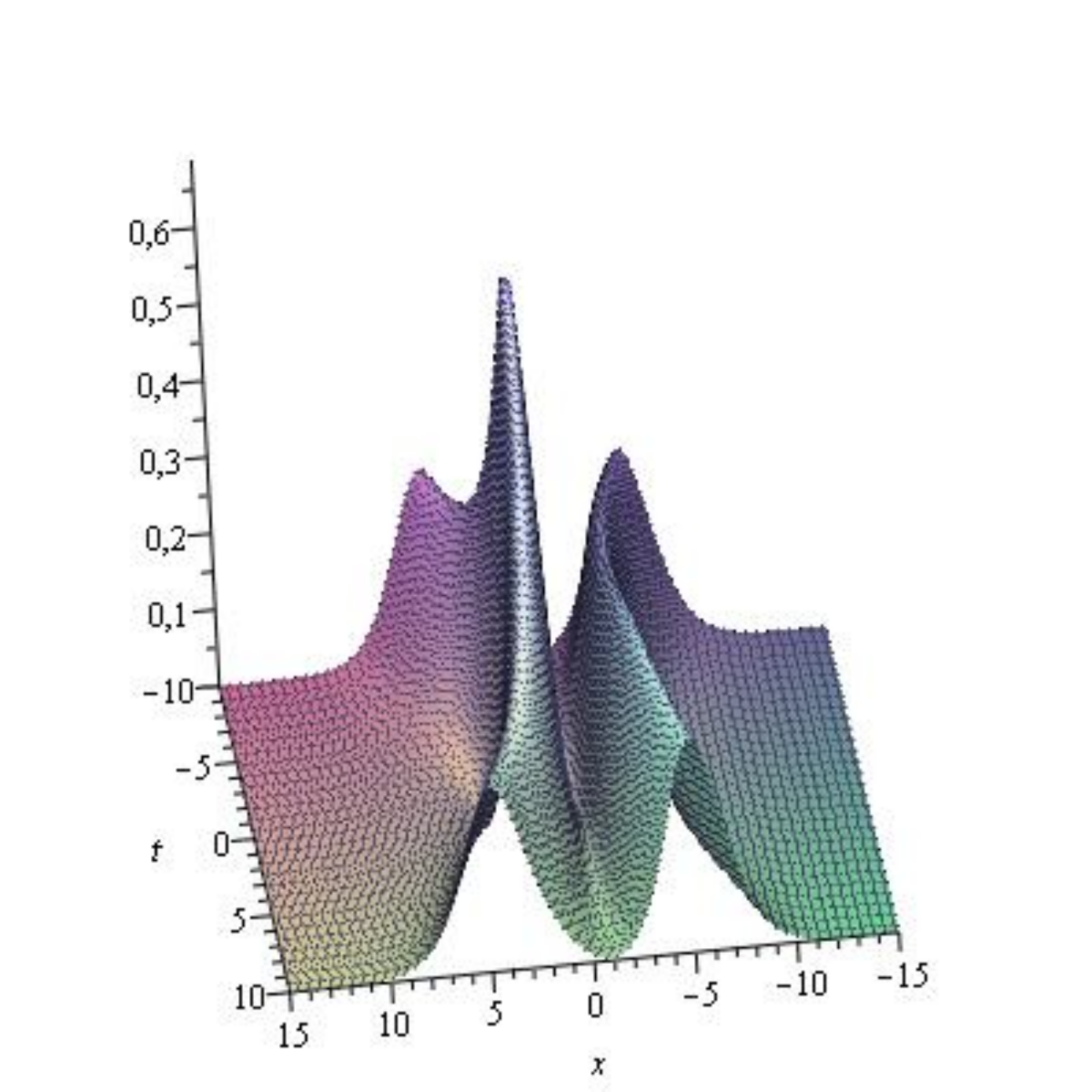}}\hfill
\subfigure[]{\label{fig:2c}\includegraphics[width=34mm]{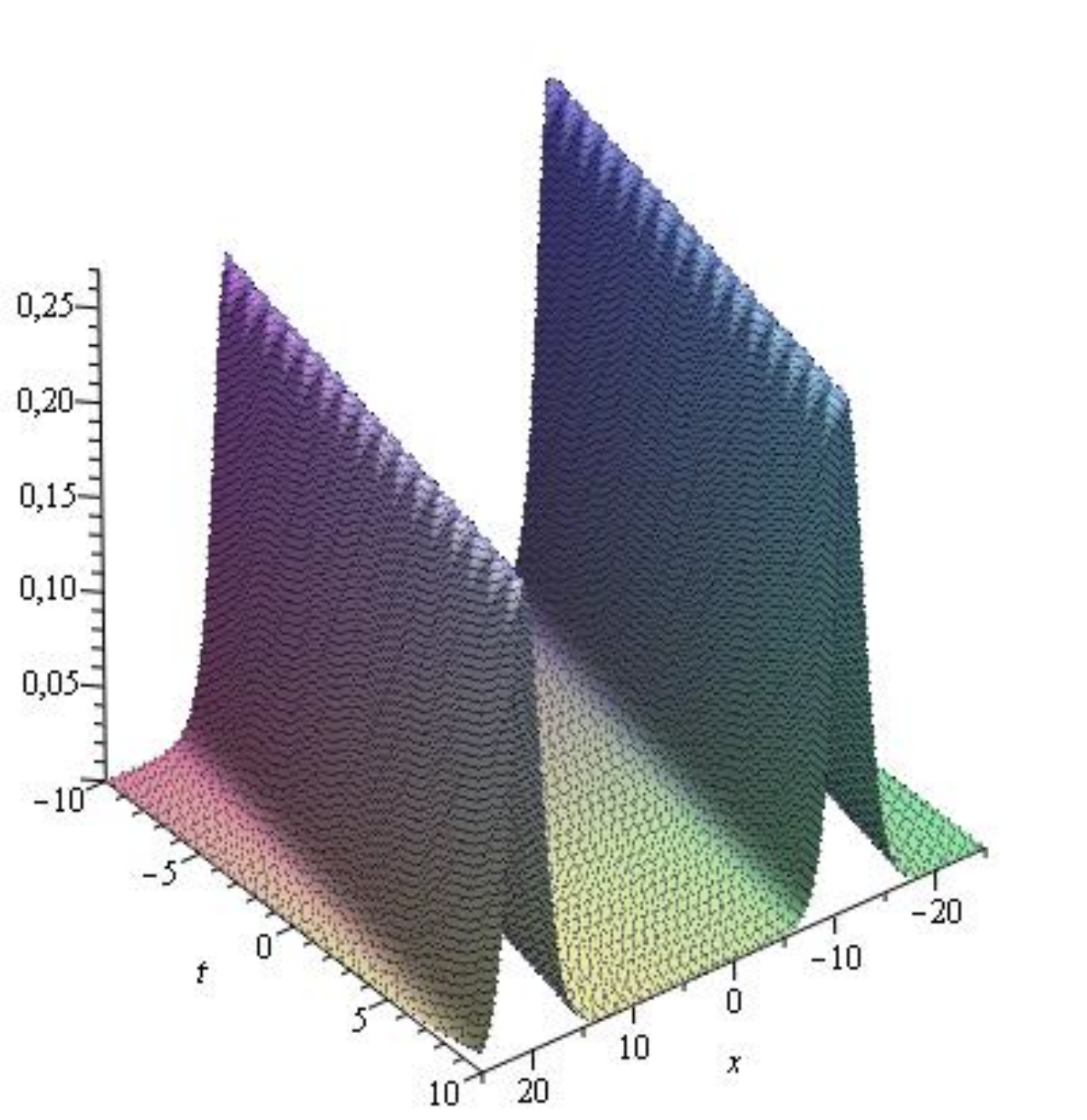}}
\caption{Different types of two-soliton solutions for the equation (\ref{denk3}).}
\end{figure}
\subsection{Three-Soliton Solution for the Standard NLS Equation}
The conditions on the parameters of three-soliton solution of the standard NLS equation (\ref{denk3}) can be easily found
by the same analysis used in Section III.A as
\begin{equation}
1)\, \bar{a}=-a,\quad 2)\, \ell_i=\bar{k}_i,\,\, i=1, 2, 3,\quad 3)\, e^{\alpha_i}=ke^{\bar{\delta}_i},\,\, i=1, 2, 3.
\end{equation}

\vspace{0.3cm}
\noindent
{\bf Example 4.}\, To illustrate some examples of three-soliton solution for the standard NLS equation we give particular values, satisfying above constraints, to the parameters of
the solution. The graphs of the functions $|q(t,x)|^2$ corresponding to
$ (k_1, l_1, k_2, l_2, k_3, l_3)=\Big(-\frac{1}{2}-\frac{2}{5}i,-\frac{1}{2}+\frac{2}{5}i, -\frac{13}{25}-\frac{2}{5}i,-\frac{13}{25}+\frac{2}{5}i,-\frac{27}{50}-\frac{2}{5}i,-\frac{27}{50}+\frac{2}{5}i\Big)$, $ (k_1, l_1, k_2, l_2, k_3, l_3)=\Big(-\frac{1}{2}-\frac{2}{5}i,-\frac{1}{2}+\frac{2}{5}i, -\frac{13}{25}-\frac{2}{5}i,-\frac{13}{25}+\frac{2}{5}i,-\frac{27}{50}+\frac{2}{5}i,-\frac{27}{50}-\frac{2}{5}i\Big)$, and $ (k_1, l_1, k_2, l_2, k_3, l_3)=\Big(-\frac{1}{2}-\frac{2}{5}i,-\frac{1}{2}+\frac{2}{5}i, -\frac{13}{25}+\frac{2}{5}i,-\frac{13}{25}-\frac{2}{5}i,\frac{27}{50}-\frac{2}{5}i,\frac{27}{50}-\frac{2}{5}i\Big)$ with $(e^{\alpha_j},e^{\delta_j},k,a)=(-1+i,1+i,-1,i)$, $j=1, 2, 3$ are given in Figures 3.a, 3.b, and 3.c, respectively.

\newpage

\begin{figure}[h!]
\centering     
\subfigure[]{\label{fig:3a}\includegraphics[width=34mm]{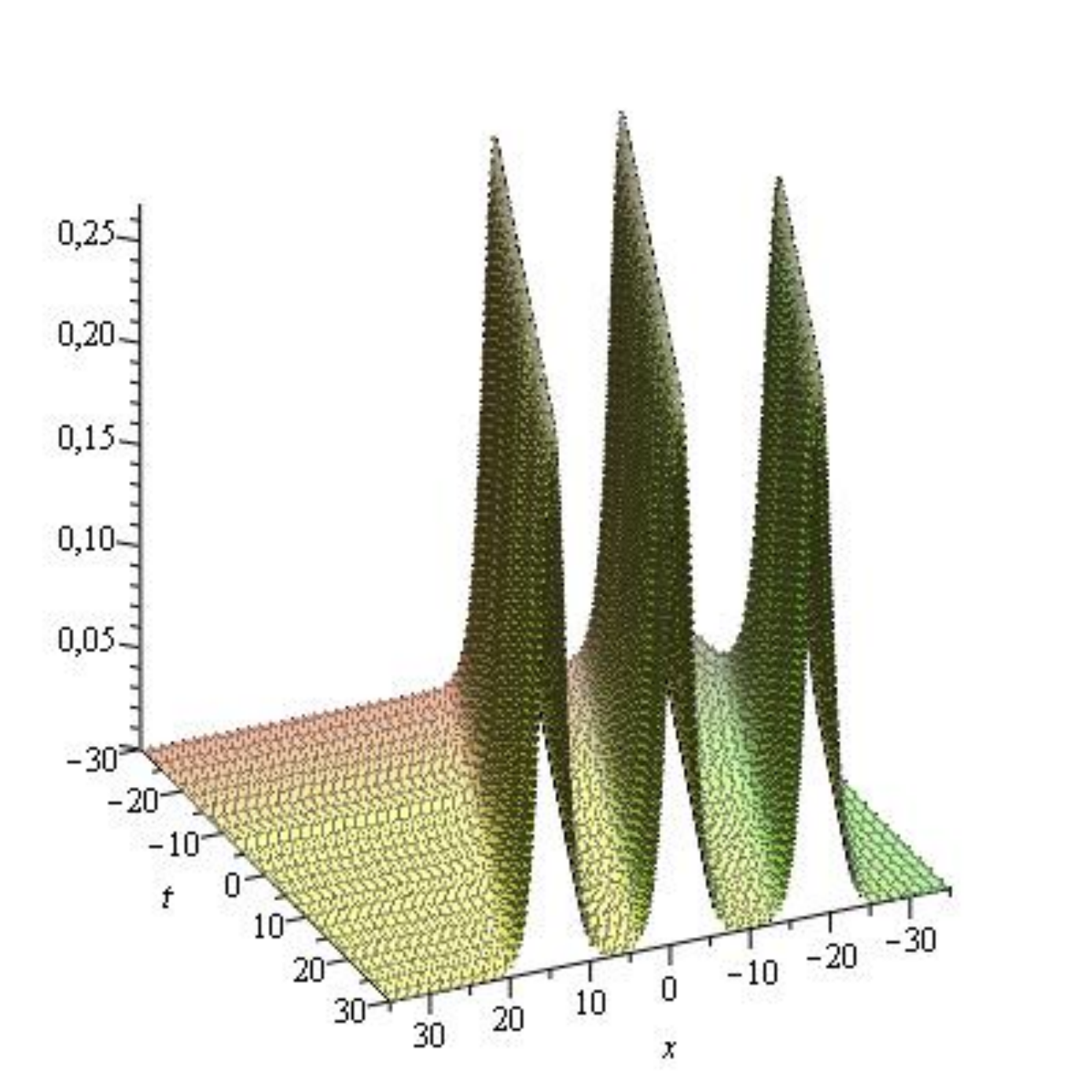}}\hfill
\subfigure[]{\label{fig:3b}\includegraphics[width=34mm]{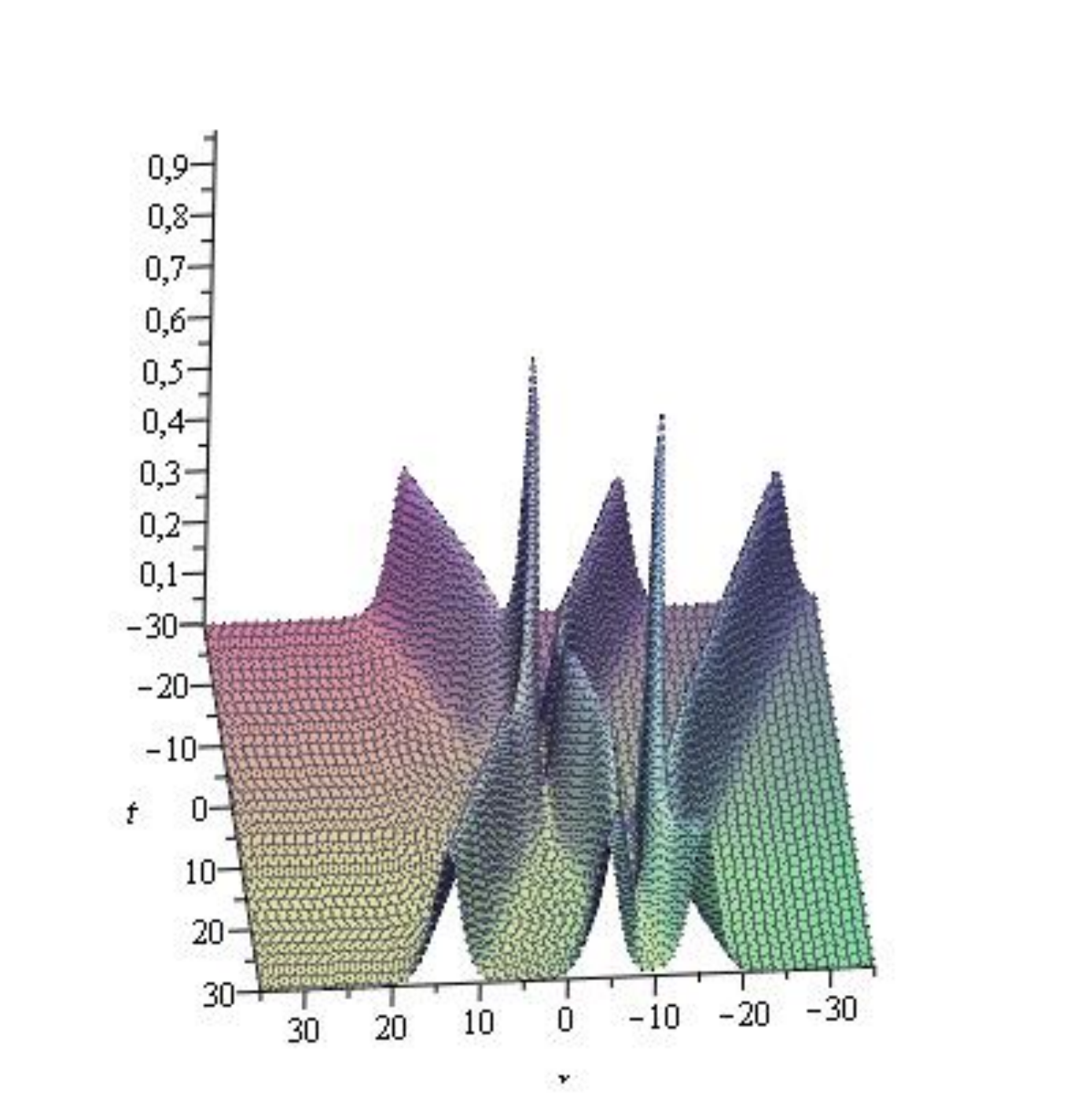}}\hfill
\subfigure[]{\label{fig:3c}\includegraphics[width=34mm]{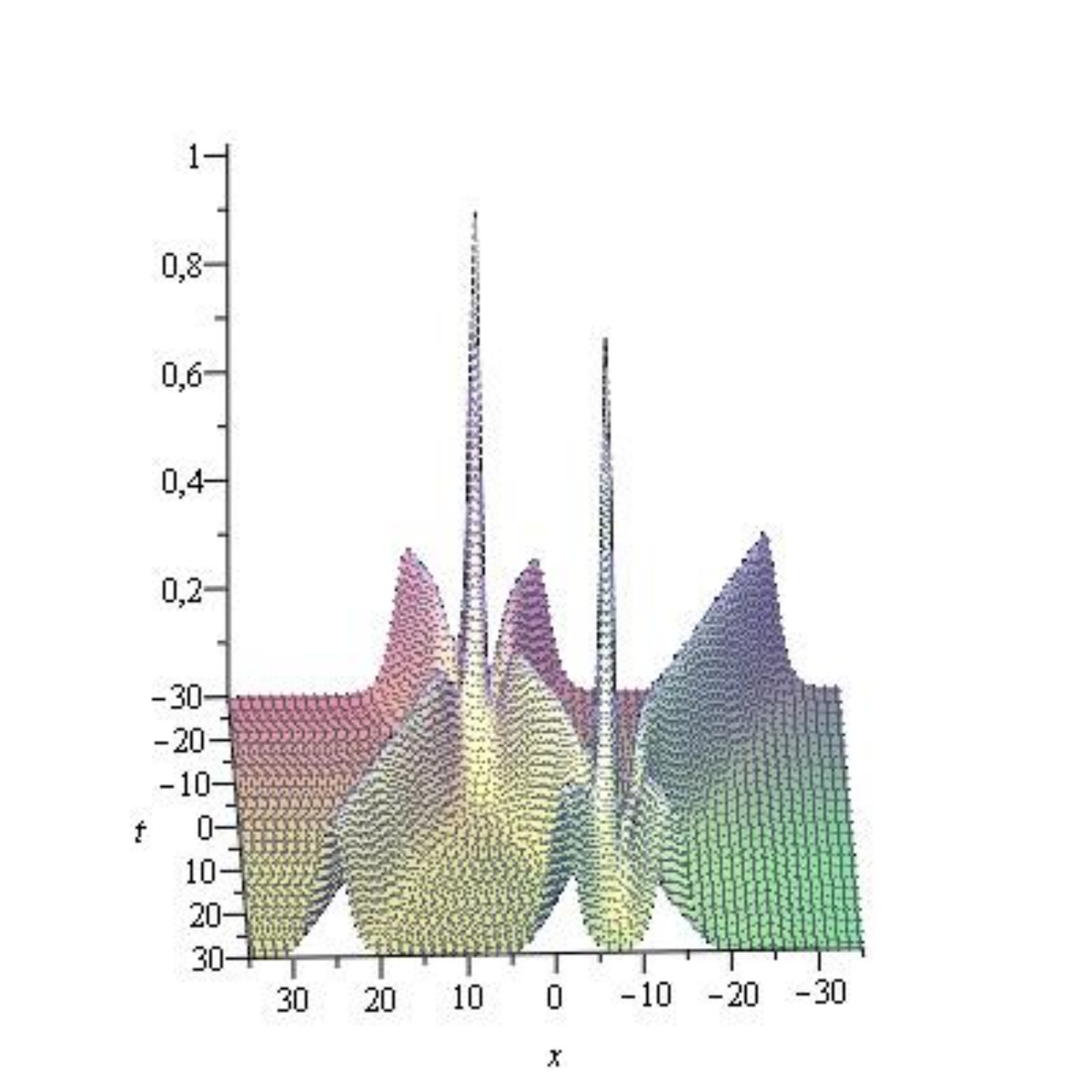}}
\caption{Different types of three-soliton solutions for the equation (\ref{denk3}).}
\end{figure}

\section{Nonlocal Reduction of the NLS System}
In this section we use the reduction (\ref{non}) introduced by Ablowitz and Musslimani \cite{AbMu1}-\cite{AbMu3} to obtain
soliton solutions for three different nonlocal NLS equations (\ref{denk41})-(\ref{denk43}) with the condition
\begin{equation}\label{nonlocalNLScond1}
\bar{a}=- \varepsilon_{1} a
\end{equation}
satisfied.

\subsection{One-Soliton Solution for Nonlocal NLS Equation}
Here we find the conditions on the parameters of one-soliton solution of the NLS system to satisfy the equality (\ref{non}).
We must have
\begin{equation}
\displaystyle \frac{e^{k_2x-\frac{k_2^2}{2a}t+\delta_2}}{1+Ae^{(k_1+k_2)x+\frac{(k_1^2-k_2^2)}{2a}t+\delta_1+\delta_2  }}
=k\frac{e^{\bar{k}_1\varepsilon_2x+\frac{\bar{k}_1^2}{2\bar{a}}\varepsilon_1t+\bar{\delta}_1}}
{1+\bar{A}e^{(\bar{k}_1+\bar{k}_2)\varepsilon_2x+\frac{(\bar{k}_1^2-\bar{k}_2^2)}{2\bar{a}}\varepsilon_1t+\bar{\delta}_1+\bar{\delta}_2}}
\end{equation}
yielding the conditions
\begin{align}\label{7conds}
\displaystyle &i)\, k_2=\varepsilon_2\bar{k}_1,\quad  ii)\,-\frac{k_2^2}{2a}=\frac{\bar{k}_1^2}{2\bar{a}}\varepsilon_1,\quad iii)\, e^{\delta_2}=ke^{\bar{\delta}_1},\quad
iv)\, \bar{A}=A,\nonumber\\
 &v)\,(k_1+k_2)=(\bar{k}_1+\bar{k}_2)\varepsilon_2,\quad
 vi)\,\frac{(k_1^2-k_2^2)}{2a}=\frac{(\bar{k}_1^2-\bar{k}_2^2)}{2\bar{a}}\varepsilon_1,\quad vii)\, e^{\delta_1+\delta_2}=e^{\bar{\delta}_1+\bar{\delta}_2}.
\end{align}
From $i)$ we have $k_2^2=\bar{k}_1^2$. If we use this relation in the left hand side of $ii)$ with (\ref{nonlocalNLScond1})
we get that the condition $ii)$ is satisfied directly since
\begin{equation*}
\displaystyle -\frac{k_2^2}{2a}=-\frac{\bar{k}_1^2}{2a}=\frac{\bar{k}_1^2}{2\bar{a}}\varepsilon_1.
\end{equation*}

\noindent For $iv)$ we only need the equality $(k_1+k_2)^2=(\bar{k}_1+\bar{k}_2)^2$ holds. Indeed it is satisfied directly since
\begin{equation*}
(k_1+k_2)^2=(\bar{k}_2\varepsilon_2+\bar{k}_1\varepsilon_2)^2=(\bar{k}_1+\bar{k}_2)^2
\end{equation*}
with the condition given in $i)$.

\noindent The condition $v)$ is already true since
\begin{equation*}
(k_1+k_2)=(\bar{k}_2\varepsilon_2+\bar{k}_1\varepsilon_2)=(\bar{k}_1+\bar{k}_2)\varepsilon_2
\end{equation*}
by the condition $k_2=\bar{k}_1\varepsilon_2$ or equivalently $k_1=\bar{k}_2\varepsilon_2$. Similarly, $vi)$
is satisfied directly since
\begin{equation*}
\displaystyle \frac{(k_1^2-k_2^2)}{2a}=\frac{(\bar{k}_2^2-\bar{k}_1^2)}{-2\varepsilon_1\bar{a}}=\frac{(\bar{k}_1^2-\bar{k}_2^2)}{2\bar{a}}\varepsilon_1,
\end{equation*}
by $k_2^2=\bar{k}_1^2$, $k_1^2=\bar{k}_2^2$, and $\bar{a}=-\varepsilon_1a$.

\noindent In Section 3.1 we proved that the condition $vii)$ is satisfied automatically by the condition $iii)$. Hence for one-soliton solutions of the nonlocal reductions of the NLS system we have obtained the following conditions:
\begin{equation}\label{NLS1sol3conds}
1)\,\bar{a}=-\varepsilon_1 a,\quad 2)\, k_2=\bar{k}_1\varepsilon_2,\quad 3)\, e^{\delta_2}=ke^{\bar{\delta}_1}.
\end{equation}
Therefore one-soliton solution of the nonlocal NLS equations is given by
\begin{equation}\label{1solq(x,t)}
\displaystyle q(t,x)=\frac{e^{k_1x+\frac{k_1^2}{2a}t+\delta_1}}{1-\frac{e^{(k_1+k_2)x+\Big(\frac{k_1^2}{2a}-\frac{k_2^2}{2a} \Big)t+\delta_1+\delta_2}}{(k_1+k_2)^2}}
\end{equation}
with the conditions (\ref{NLS1sol3conds}) satisfied.

\noindent Now and then we will consider only the case $(\varepsilon_1,\varepsilon_2)=(1,-1)$ (S-symmetric case).
Here the nonlocal reduction is $r(t,x)=k\bar{q}(t,-x)$ giving $\bar{a}=-a$, $k_2=-\bar{k}_1$, and
\begin{equation}\label{casebonesolnNLS}
\displaystyle aq_t(t,x)=\frac{1}{2}q_{xx}(t,x)-kq(t,x)\bar{q}(t,-x)q(t,x),
\end{equation}
with $e^{\delta_2}=ke^{\bar{\delta}_1}$. From $\bar{a}=-a$, we have $a=iy$, $y\in \mathbb{R}$. If $k_1=\alpha+i\beta$, $\alpha, \beta \in \mathbb{R}$ then the solution of (\ref{casebonesolnNLS}) becomes
\begin{equation}
\displaystyle q(t,x)=\frac{e^{(\alpha+i\beta)x+\frac{(\alpha+i\beta)^2}{2iy}t+\delta_1}}{1+k\frac{e^{2i\beta x+\frac{2\alpha\beta}{y}t+\delta_1+\bar{\delta}_1}}{4\beta^2}},
\end{equation}
where $\beta\neq 0$. Here the solution is complex valued. Hence let us consider the real valued function $|q(t,x)|^2$.
We have
\begin{equation}\label{casebmoduleonesolnNLS}
\displaystyle |q(t,x)|^2=\frac{16\beta^4e^{2\alpha x+\frac{2\alpha\beta}{y}t+\delta_1+\bar{\delta}_1}}{(ke^{\frac{2\alpha\beta}{y}t+\delta_1+\bar{\delta}_1} +4\beta^2\cos(2\beta x))^2+16\beta^4\sin^2(2\beta x)}.
\end{equation}
This function is singular at $\displaystyle x=\frac{n \pi}{2 \beta}$, $ke^{\frac{2\alpha\beta}{y}t+\delta_1+\bar{\delta}_1} +4\beta^2\, (-1)^n=0$ both for focusing and defocusing cases. If $\alpha=0$, the function (\ref{casebmoduleonesolnNLS}) becomes
\begin{equation}\label{casebmoduleonesolnNLSalpha=0}\displaystyle
|q(t,x)|^2=\frac{2\beta^2}{k[B+\cos(2\beta x)]},
\end{equation}
for $\displaystyle B=\frac{\rho^2+16\beta^4}{8\rho\beta^2}$ where $\rho=ke^{\delta_1+\bar{\delta}_1}$. Clearly, the solution (\ref{casebmoduleonesolnNLSalpha=0})
is non-singular if $B>1$ or $B<-1$.\\

 \vspace{0.3cm}
\noindent
{\bf Example 5.}\,  For the set of parameters $ (k_1, k_2, e^{\delta_1}, e^{\delta_2}, k, a)=(i, i, i, -i, 1, \frac{i}{2})$, we get the solution
\begin{equation}
  \displaystyle q(t,x)=\frac{4ie^{ix+it}}{4+e^{2ix}},
\end{equation}
and therefore
\begin{equation}\label{casebNLSex1}
\displaystyle |q(t,x)|^2=\frac{16}{17+8\cos(2x)}.
\end{equation}
This solution represents a periodic solution. Its graph is given in Figure 4.\\

\vspace{0.3cm}
\noindent
{\bf Example 6.}\,  In addition to the solution given with the conditions (\ref{NLS1sol3conds}) we have  another possible solution of $r(t,x)=k \bar{q}(t,-x)$ which is given by
\begin{equation}
q(t,x)=e^{\frac{k_{1}^2}{2a}t+\delta_{1}}\, \frac{e^{k_{1} x}}{1+e^{2 k_{1} x}},
\end{equation}
where $e^{\delta_{2}}=k e^{\bar{\delta}_{1}}$, $A k e^{\bar{\delta}_{1}+\delta_{1}}=1$, $k_{2}=k_{1}$, and $k_{1}$ is real. Here $\bar{a}=-a$. Hence
\begin{equation}
|q(t,x)|^2=-\frac{k_{1}^2}{k} \mathrm{sech}^2{(k_{1} x)},
\end{equation}
which represents a stationary soliton solution for the focusing case ($k<0$). For example, if we consider $\displaystyle k_1=\frac{1}{2}$ and $e^{\delta_1}=1+i$ giving
$\displaystyle k=-\frac{1}{2}<0$, the above function becomes
\begin{equation}\label{casebNLSex2}\displaystyle
|q(t,x)|^2=\frac{1}{2}\mathrm{sech}^2\Big(\frac{1}{2}x\Big),
\end{equation}
which represents a soliton. Its graph is given in Figure 5.
\begin{center}
\begin{figure}[ht]
\centering
\begin{minipage}{0.4\linewidth}
\centering
\includegraphics[angle=0,scale=.21]{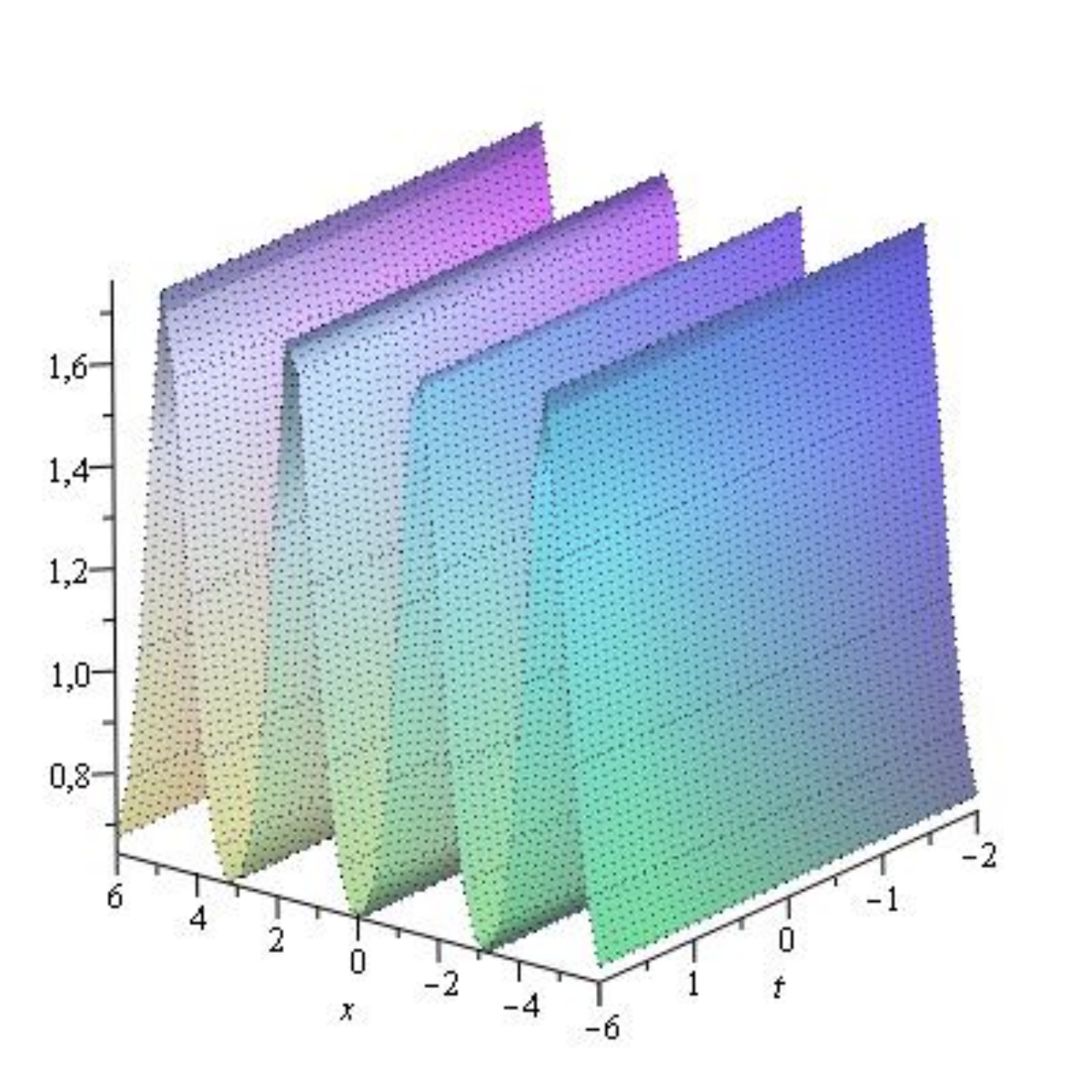}
\caption{Periodic solution for (\ref{casebNLSex1}).}
\end{minipage}
\hfill
\begin{minipage}{0.45\linewidth}
\centering
\includegraphics[angle=0,scale=.21]{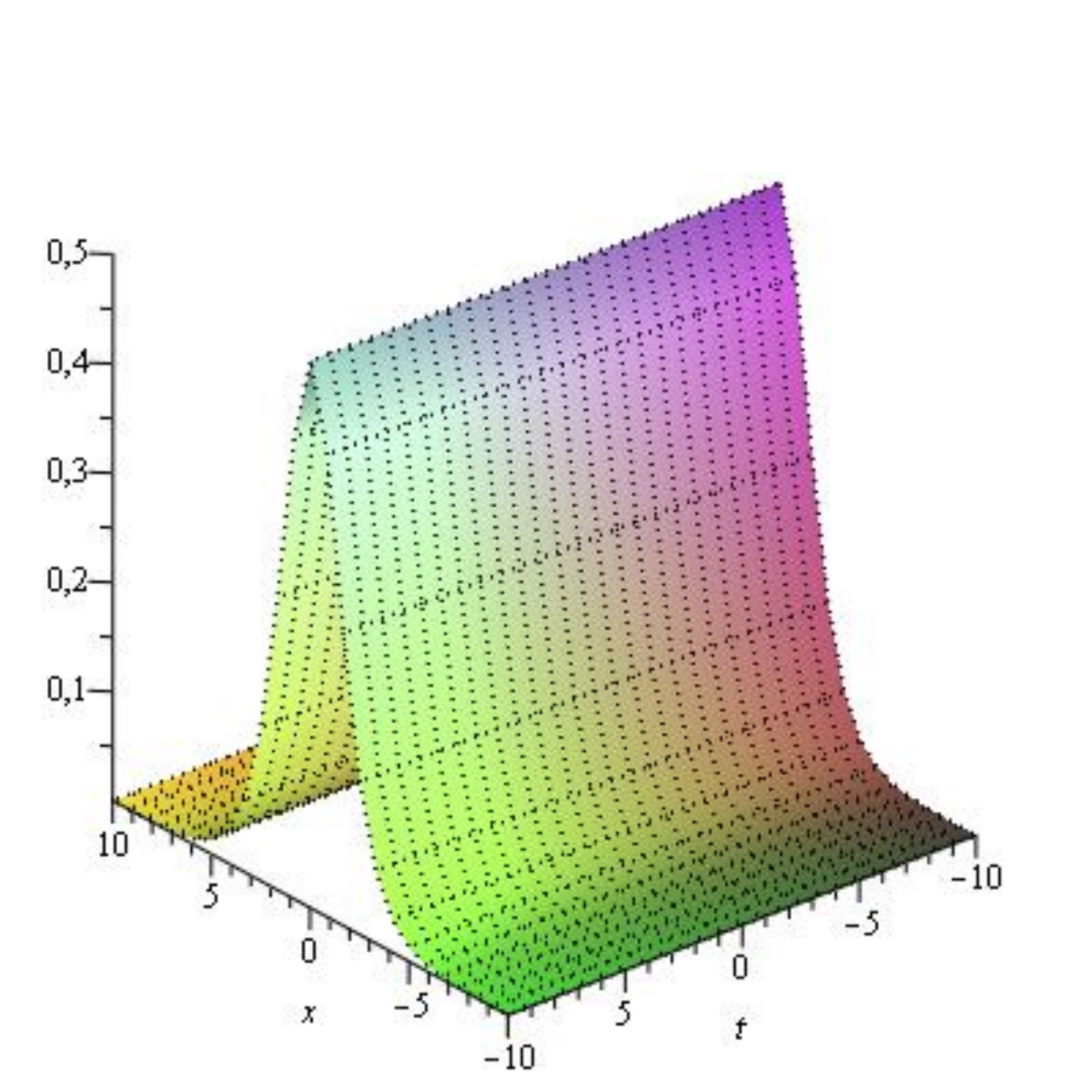}
\caption{One-soliton solution for (\ref{casebNLSex2}). }
\end{minipage}
\end{figure}
\end{center}

\noindent \textbf{Remark 2.} In Ref. \cite{SSL}, the authors studied a particular form of $S$-symmetric nonlocal NLS equation (\ref{denk42}) where $\displaystyle a=\frac{i}{2}$ and $k=-1$,
\begin{equation}\label{ssleqn}
iq_t(t,x)=q_{xx}(t,x)+2q(t,x)q^*(t,-x)q(t,x).
\end{equation}
Here * is used for complex conjugation.  In \cite{SSL}, one-soliton solution
of the nonlocal equation (\ref{ssleqn}) is given as
\begin{equation}\label{sslonesol}\displaystyle
q(t,x)=\frac{\alpha_1e^{i\bar{\ell}_1x+i\bar{\ell}_1^2t+{\bar{\xi}_1}^{(0)}}}
{1-\frac{\alpha_1\beta_1}{(\ell_1+\bar{\ell}_1)^2}e^{i(\bar{\ell}_1+\ell_1)x+i(\bar{\ell}_1^2-\ell_1^2)t+{\bar{\xi}_1}^{(0)}+{\xi_1}^{(0)}  }}.
\end{equation}
Here we expressed their parameters $k_1$, $\bar{k}_1$ of \cite{SSL} as $\ell_1$, $\bar{\ell}_1$ respectively,
not to mix with our $k_1$, $k_2$. Under the conditions $\displaystyle a=\frac{i}{2}$, $k=-1$, $e^{\delta_1}=\alpha_1e^{\bar{\xi}_1^{(0)}}$, and $e^{\bar{\delta}_1}=\beta_1e^{\xi_1^{(0)}}$, the solution (\ref{sslonesol}) becomes equivalent to our case. They also give the function $q^*(t,-x)$ as
\begin{equation}\label{ssloneconjugate}\displaystyle
q^{*}(t,-x)=\frac{\beta_1e^{i\ell_1x-i\ell_1^2t+{\xi_1}^{(0)}}}{1-\frac{\alpha_1\beta_1}
{(\ell_1+\bar{\ell}_1)^2}e^{i(\bar{\ell}_1+l_1)x+i(\bar{\ell}_1^2-\ell_1^2)t+{\bar{\xi}_1}^{(0)}+{\xi_1}^{(0)}  }}
\end{equation}
and define the constants $\ell_1, \bar{\ell}_1, \alpha_1, \beta_1, \xi_1^{(0)}$, and $\bar{\xi}_1^{(0)}$ as arbitrary
complex constants. But obviously from the relation between the functions $q(t,x)$ and $q^*(t,-x)$ the following constraints must be satisfied
\begin{equation}\label{const11}
\alpha_1^{*}=\beta_1,\quad \ell_1=(\bar{\ell}_1)^{*},\quad  \xi_1^{(0)}=(\bar{\xi}_1^{(0)})^{*}.
\end{equation}
These conditions are equivalent to our conditions coming from the reduction (\ref{non}) for the S-symmetric case which were missed in \cite{SSL}. Because of this fact the example given in \cite{SSL}  with the parameters chosen as $\ell_1=0.4+i$, $\bar{\ell}_1=-0.4+i$, $\alpha_1=1+i$, and
$\beta_1=1-i$ is not valid. They claim that they find the non-singular most general one-bright soliton solution of the equation (\ref{ssleqn}) which is not correct, because the above constraints (\ref{const11}) are not satisfied by the parameters they have choosen. Indeed such specific parameters they use are not allowed, since $\ell_1=0.4+i\neq -0.4-i=(\bar{\ell}_1)^{*}$. Note that if we use the parameters not satisfying (\ref{const11}) that they give and e.g. $e^{{\bar{\xi}_1}^{(0)}}=1+i$ and $e^{{\xi_1}^{(0)}}=-1+i$  in the solution then the solution (\ref{sslonesol}) and $q^*(t,-x)$ becomes
\begin{equation}\displaystyle
q(t,x)=\frac{2ie^{(-1-\frac{2}{5}i)x+(\frac{4}{5}-\frac{21}{25}i)t}}{1-e^{-2x+\frac{8}{5}t}}, \quad q^*(t,-x)=\frac{-2ie^{(1-\frac{2}{5}i)x+(\frac{4}{5}+\frac{21}{25}i)t}}{1-e^{2x+\frac{8}{5}t}}.
\end{equation}
One can easily check that the nonlocal NLS equation (\ref{ssleqn}) is not satisfied by the above functions.

If we take the parameters satisfying (\ref{const11}), for instance
$\ell_1=0.4+i$, $\bar{\ell}_1=0.4-i$, $\alpha_1=1+i$, and
$\beta_1=1-i$ with $ \xi_1^{(0)}=\bar{\xi}_1^{(0)}=0$, then the solution (\ref{sslonesol}) becomes
\begin{equation}\displaystyle
  q(t,x)=\frac{(1+i)e^{(1+\frac{2}{5}i)x+(\frac{4}{5}-\frac{21}{25}i)t}}{1-\frac{25}{8}e^{\frac{4}{5}ix+\frac{8}{5}t}},
\end{equation}
and so
\begin{equation}\displaystyle
|q(t,x)|^2=\frac{2e^{2x+8t}}{(\frac{25}{8}e^{\frac{8}{5}t}-\cos(\frac{4}{5}x))^2+\sin^2(\frac{4}{5}x)},
\end{equation}
which is not a solitary wave. Indeed it has singularity at $ (x,t)=\Big(\frac{5}{2}n\pi,\frac{5}{8}\ln(\frac{8}{25})\Big)$, $n$ is an integer.

We understand that  the authors of Ref. \cite{SSL} are solving the NLS system of equations (\ref{denk1}) and (\ref{denk2}) rather then solving nonlocal NLS equation (\ref{denk42}) as they claim. They treat $q^*(t,-x)$ as a separate quantity than $q(t,x)$ rather than using the equivalence $q^*(t,-x)=(q(t,x))^*|_{x \to -x}$. That is the reason why they miss the constraint equations (\ref{const11}) for the parameters of the one-soliton solution.

\subsection{Two-Soliton Solution for Nonlocal NLS Equation}
We obtain the conditions on the parameters of two-soliton solution of the NLS system to satisfy the equality (\ref{non}), where
the function $r(t,x)$ is given in (\ref{NLS2solr(t,x)}) and $k\bar{q}(\varepsilon_1t,\varepsilon_2x)$ is
\begin{equation}
\displaystyle k\bar{q}(\varepsilon_1t,\varepsilon_2x)=k\frac{e^{\bar{\theta}_1}+e^{\bar{\theta}_2}
+\bar{A}_1e^{\bar{\theta}_1+\bar{\theta}_2+\bar{\eta}_1}+\bar{A}_2e^{\bar{\theta}_1+\bar{\theta}_2+\bar{\eta}_2}}
{1+e^{\bar{\theta}_1+\bar{\eta}_1+\bar{\alpha}_{11}}+e^{\bar{\theta}_1+\bar{\eta}_2+\bar{\alpha}_{12}}+e^{\bar{\theta}_2+\bar{\eta}_1+\bar{\alpha}_{21}}
+e^{\bar{\theta}_2+\bar{\eta}_2+\bar{\alpha}_{22}}+\bar{M}e^{\bar{\theta}_1+\bar{\theta}_2+\bar{\eta}_1+\bar{\eta}_2}},
\end{equation}
where
\begin{align*}\displaystyle
& \bar{\theta}_i=\varepsilon_2\bar{k}_ix+\varepsilon_1\frac{\bar{k}_i^2}{2\bar{a}}t+\bar{\delta}_i, \\
& \bar{\eta}_i=\varepsilon_2\bar{\ell}_ix-\varepsilon_1\frac{\bar{\ell}_i^2}{2\bar{a}}t+\bar{\alpha}_i,
\end{align*}
for $i=1,2$.
Here we have the following conditions that must be satisfied:
\begin{align}
&i)\, e^{{\eta}_i}=ke^{\bar{\theta}_i}, i=1, 2,\quad ii)\, e^{\theta_1+\eta_1+\eta_2}=ke^{\bar{\theta}_1+\bar{\theta}_2+\bar{\eta}_1},\quad
iii)\, B_i=\bar{A}_i, i=1, 2,\nonumber\\
&iv)\, e^{\theta_2+\eta_1+\eta_2}=ke^{\bar{\theta}_1+\bar{\theta}_2+\bar{\eta}_2}, \quad v)\, e^{\theta_1+\eta_1}=e^{\bar{\theta}_1+\bar{\eta}_1}, \quad
vi)\, e^{\theta_1+\eta_2}=e^{\bar{\theta}_2+\bar{\eta}_1},\nonumber\\
&vii)\, e^{\theta_2+\eta_1}=e^{\bar{\theta}_1+\bar{\eta}_2},\quad viii)\, e^{\theta_2+\eta_2}=e^{\bar{\theta}_2+\bar{\eta}_2},\quad ix) \, e^{{\alpha}_{ij}}=e^{\bar{\alpha}_{ji}}, i, j=1, 2,\nonumber\\
&x)\, M=\bar{M},\quad xi)\, e^{\theta_1+\theta_2+\eta_1+\eta_2}=e^{\bar{\theta}_1+\bar{\theta}_2+\bar{\eta}_1+\bar{\eta}_2}.\nonumber\\
&
\end{align}
From the condition $i)$ we get
\begin{equation}
\displaystyle \ell_ix-\frac{\ell_i^2}{2a}t=\varepsilon_2\bar{k}_ix+\varepsilon_1\frac{\bar{k}_i^2}{2\bar{a}}t,\quad e^{{\alpha}_i}=ke^{\bar{\delta}_i}, i=1, 2,
\end{equation}
yielding $\ell_i=\varepsilon_2\bar{k}_i$, $i=1, 2$. The coefficients of $t$ in the above equality are directly equal with this relation and $\bar{a}=-\varepsilon_1a$ that we have obtained previously. All the other conditions $ii)$-$xi)$ are also satisfied automatically by the following conditions:
\begin{equation}
1)\, \bar{a}=-\varepsilon_1 a,\quad 2)\, \ell_i=\varepsilon_2\bar{k}_i, i=1, 2,\quad 3)\,e^{\alpha_i}=ke^{\bar{\delta}_i}, i=1, 2.
\end{equation}

\noindent For particular choice of the parameters let us present some solutions of the nonlocal reduction of the NLS system only for $(\varepsilon_1,\varepsilon_2)=(1,-1)$ (S-symmetric case).
In this case we have $\bar{a}=-a$, $\ell_i=-\bar{k}_i$, and $e^{\alpha_i}=ke^{\bar{\delta}_i}$ for $i=1, 2$.

\vspace{0.3cm}
\noindent
{\bf Example 7.}\, Consider the set of the parameters $\displaystyle (k_1,\ell_1,k_2,\ell_2)=(\frac{i}{4},\frac{i}{4},i,i)$ with $\displaystyle (e^{\alpha_j},e^{\delta_j},k,a)=(1,1,1,\frac{i}{2})$ for $j=1, 2$. The solution $q(t,x)$ becomes
\begin{equation}
\displaystyle
q(t,x)=\frac{e^{\frac{1}{4}ix+\frac{1}{16}it}+e^{ix+it}+\frac{36}{25}e^{\frac{3}{2}ix+it}+\frac{9}{100}e^{\frac{9}{4}ix+\frac{1}{16}it}}
{1+4e^{\frac{1}{2}ix}+\frac{16}{25}e^{\frac{5}{4}ix-\frac{15}{16}it}+\frac{16}{25}e^{\frac{5}{4}ix+\frac{15}{16}it}+\frac{1}{4}e^{2ix}
+\frac{81}{625}e^{\frac{5}{2}ix}}
\end{equation}
and so the function $|q(t,x)|^2$ is
\begin{equation}\label{casebNLStwoex1}   \displaystyle
|q(t,x)|^2=\frac{Y_1}{Y_2},
\end{equation}
where
\begin{align*}
&Y_1=625(20000\cos\Big(\frac{3}{4}x+\frac{15}{16}t\Big)+28800\cos\Big(\frac{5}{4}x+\frac{15}{16}t\Big)
+2592\cos\Big(-\frac{3}{4}x+\frac{15}{16}t\Big)\\&+1800\cos\Big(-\frac{5}{4}x+\frac{15}{16}t\Big)
+28800\cos\Big(\frac{1}{2}x\Big)+1800\cos 2x+40817),
\end{align*}
and
\begin{align*}
&Y_2=100\Big(340000\cos\Big(\frac{3}{4}x+\frac{15}{16}t\Big)+90368\cos\Big(\frac{5}{4}x+\frac{15}{16}t\Big)+340000\cos\Big(-\frac{3}{4}x+\frac{15}{16}t\Big)
\\&+90368\cos\Big(-\frac{5}{4}x+\frac{15}{16}t\Big)+504050\cos\Big(\frac{1}{2}x\Big)+125000\cos\Big(\frac{3}{2}x\Big)+16200\cos\Big(\frac{5}{2}x\Big)\\
&+96050\cos 2x+51200\cos\Big(\frac{15}{8}t\Big)\Big)+111865601.
\end{align*}
The graph of (\ref{casebNLStwoex1}) is given in Figure 6.
\begin{center}
\begin{figure}[h]
\begin{minipage}{1\textwidth}
\centering
\includegraphics[angle=0,scale=.30]{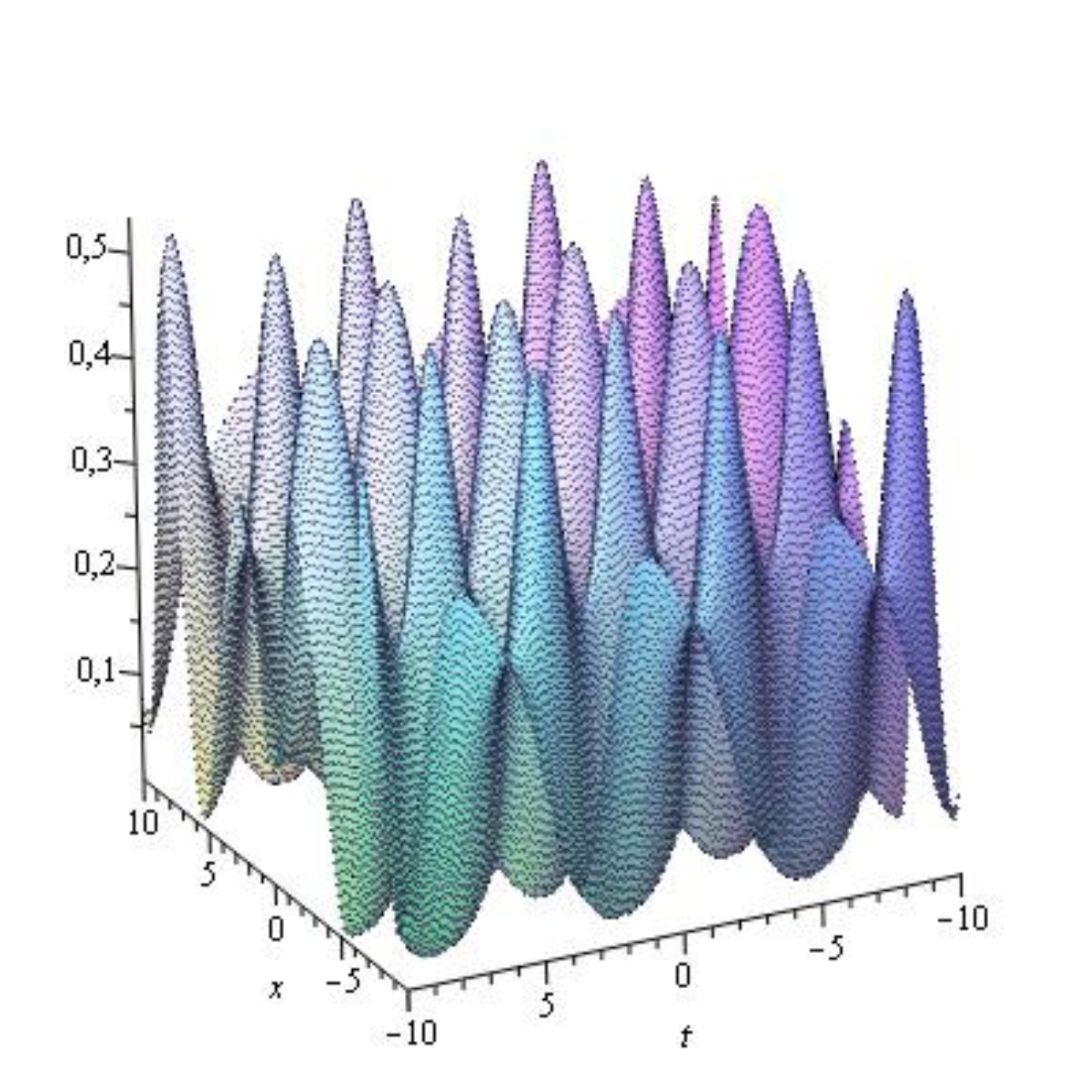}
\caption{Breather type of wave solution for (\ref{casebNLStwoex1}).}
\end{minipage}
\end{figure}
\end{center}

\noindent \textbf{Remark 3.} Two-soliton solution presented in Ref. \cite{SSL} has  the same flaw as stated in Remark 2. They chose the parameters of their solution not satisfying the constraint equations. Because of the relation between the functions
$q(t,x)$ and $q^*(t,-x)$ their parameters must satisfy the following constraints,
\begin{equation}
1)\,\alpha_p^*=\beta_p,\quad  2)\,\ell_p=(\bar{\ell}_p)^*, p=1, 2,\quad 3)\,e^{\gamma_j}=(e^{\Delta_j})^*,
\end{equation}
where $j=\{1,2,3,4,11,12,21,22,23,24,25,26,31,32\}$. Remember that we use $\ell$ and $\bar{\ell}$ instead of the parameters $k$ and $\bar{k}$  (parameters of \cite{SSL}) respectively.
However, they have taken the parameters as in the form $\bar{\ell}_1=a_1+b_1i$, $\ell_1=-a_1+b_1i$, $\bar{\ell}_2=c_1+d_1i$, and $\ell_2=-c_1+d_1i$
for some specific values of $a_p$, $b_p$, $c_p$, and $d_p$, $p=1, 2$. Clearly, the parameters do not satisfy the above constraints, hence two-soliton solution of \cite{SSL} does not satisfy the nonlocal nonlinear Schr\"{o}dinger equation (\ref{ssleqn}).

\subsection{Three-Soliton Solution for Nonlocal NLS Equation}
Similar to one- and two-soliton solution for nonlocal NLS equations, we first obtain the conditions on the parameters of
three-soliton solution of the NLS system to satisfy the equality (\ref{non}) where $r(t,x)$ is given by (\ref{NLS3solr(t,x)}) and
\begin{equation}
\displaystyle k\bar{q}(\varepsilon_1t,\varepsilon_2x)=k\frac{e^{\bar{\theta}_1}+e^{\bar{\theta}_2}+e^{\bar{\theta}_3}+\sum_{\substack{1\leq i,j,s\leq 3 \\ i<j}} \bar{A}_{ijs}e^{\bar{\theta}_i+\bar{\theta}_j+\bar{\eta}_s}
+\sum_{\substack{1\leq i,j\leq 3 \\ i<j}}\bar{V}_{ij}e^{\bar{\theta}_1+\bar{\theta}_2+\bar{\theta}_3+\bar{\eta}_i+\bar{\eta}_j}}{1+\sum_{1\leq i,j\leq3}e^{\bar{\theta}_i+\bar{\eta}_j+\bar{\alpha}_{ij}}+\sum_{\substack{1\leq i<j\leq 3 \\ 1\leq p<r \leq3}} \bar{M}_{ijpr}e^{\bar{\theta}_i+\bar{\theta}_j+\bar{\eta}_p+\bar{\eta}_r}+\bar{H}e^{\bar{\theta}_1+\bar{\theta}_2+\bar{\theta}_3+\bar{\eta}_1+\bar{\eta}_2+\bar{\eta}_3}},
\end{equation}
where
\begin{align*}\displaystyle
& \bar{\theta}_i=\varepsilon_2\bar{k}_ix+\varepsilon_1\frac{\bar{k}_i^2}{2\bar{a}}t+\bar{\delta}_i, \,i=1, 2, 3\\
& \bar{\eta}_i=\varepsilon_2\bar{\ell}_ix-\varepsilon_1\frac{\bar{\ell}_i^2}{2\bar{a}}t+\bar{\alpha}_i, \,i=1, 2, 3.
\end{align*}
Here we obtain that (\ref{non}) is satisfied by the following conditions:
\begin{equation}
1)\, \bar{a}=-\varepsilon_1 a,\quad 2)\, \ell_i=\varepsilon_2\bar{k}_i, i=1, 2, 3\quad 3)\,e^{\alpha_i}=ke^{\bar{\delta}_i}, i=1, 2, 3.
\end{equation}

\noindent For $(\varepsilon_1,\varepsilon_2)=(1,-1)$ (S-symmetric case), the constraints are $\bar{a}=-a$, $\ell_i=-\bar{k}_i$, and $e^{\alpha_i}=ke^{\bar{\delta}_i}$ for $i=1, 2, 3$. Examples of bounded and non-singular three-soliton solutions are under investigation.

\section{Conclusion}
In this work, by using the standard Hirota method, we found one-, two-, and three-soliton solutions of the integrable coupled NLS system. Then we have studied the standard and nonlocal (Ablowitz-Musslimani type) reductions of NLS system and obtained integrable time T-, space S-, and space-time ST- reversal symmetric nonlocal NLS equations. By using the reduction formulas on the soliton solutions of the coupled NLS
system we obtained one-, two-, and three-soliton solutions of the nonlocal NLS equations. It is important to note that to obtain these soliton solutions of the nonlocal NLS equations the parameters of the soliton solutions of NLS system must satisfy certain constraints for each type of nonlocal NLS equations. These constraints play critical role to obtain the soliton solutions of the nonlocal NLS equations. Although we found solutions of all types of nonlocal NLS equations we gave only the solutions of the S-symmetric case. Furthermore, we gave particular values to the parameters (satisfying the constraint equations) of the solutions and plot the graphs of $|q(t,x)|^2$ to illustrate the solutions.

\section{Acknowledgment}
  This work is partially supported by the Scientific
and Technological Research Council of Turkey (T\"{U}B\.{I}TAK).\\

\end{document}